\documentclass[
pra%
,preprint%
,amssymb, nobibnotes, aps, superscriptaddress, floatfix]{revtex4}

\usepackage{amsmath}
\usepackage{enumitem}
\usepackage{euscript}
\usepackage{graphicx}
\usepackage{amsfonts}
\usepackage{bm}

\newcommand{\fig}{Fig.\,}

\renewcommand{\imath}{\mathrm{i}}
\newcommand{\dd}{\mathrm{d}}
\newcommand{\rv}{\mathbf{r}}
\newcommand{\av}{\boldsymbol{\alpha}}
\newcommand{\as}{\alpha}
\newcommand{\kv}{\mathbf{k}}

\newcommand{\CAH}{$\mathcal{H}$}
\newcommand{\CAHt}{$\mathcal{H}(t)$}

\newcommand{\khsub}{\mathrm{KH}}
\newcommand{\phase}{\phi}

\newcommand{\psit}{\widetilde{\psi}}
\newcommand{\Vt}{\widetilde{V}}
\newcommand{\Vb}{\overline{V}}

\newcommand{\Tmat}{\mathbf{T}}
\newcommand{\Vmat}{\mathbf{V}}
\newcommand{\Pvec}{\boldsymbol{\psi}}
\newcommand{\Cvec}{\mathbf{c}}

\newcommand{\KNORM}{\frac{1}{(2 \pi)^3}}
\newcommand{\WNORM}{\frac{1}{T_\omega}}

\newcommand{\intK}{\int \dd^3 k}
\newcommand{\dk}{\Delta k}

\newcommand{\Vidx}[1]{\Vt_{#1}}
\newcommand{\Vmk}{\Vidx{m}(\kv, t)} 
\newcommand{\Vmkp}{\Vidx{m'}(\kv-\kv', t)} 
\newcommand{\Vmpkp}{\Vidx{m-m'}(\kv-\kv', t)}
\newcommand{\Vk}{\Vt(\kv)} 
\newcommand{\Vkp}{\Vt(\kv-\kv')} 
\newcommand{\Vkpo}{\Vidx{0}(\kv-\kv', t)} 
\newcommand{\Vko}{\Vidx{0}(\kv, t)} 

\newcommand{\Pidx}[1]{\psit_{#1}}
\newcommand{\Pmk}{\Pidx{m}(\kv, t)} 
\newcommand{\Pmkp}{\Pidx{m}(\kv', t)}
\newcommand{\Pmpkp}{\Pidx{m'}(\kv', t)} 

\newcommand{\Vdk}{\Vidx{\kv}}
\newcommand{\Vdkp}{\Vidx{\kv-\kv'}}

\newcommand{\Pdmk}{\Pidx{\kv m}(t)}
\newcommand{\Pdmkp}{\Pidx{\kv' m'}(t)}

\def\e#1{\mathrm{e}^{#1}}

\usepackage{color}

\usepackage{color}

\newcommand{\edit}[1]{\textcolor{black}{#1}}
\newcommand{\editt}[1]{\textcolor{black}{#1}}

\newcommand{\hide}[1]{}

\newcommand{\domark}{%
	\vbox to 0pt{
		\kern-\dp\strutbox
		\smash{\llap{*\kern1em}}
		\vss
	}%
}

\begin{document}

\title[]{Floquet approach for dynamics in short and intense laser pulses}

\author{Lukas Medi\v sauskas}
\author{Ulf Saalmann}
\author{Jan-Michael Rost }
\affiliation{Max Planck Institute for the Physics of Complex Systems, N\"{o}thnitzer Stra{\ss}e 38, D-01187 Dresden, Germany}

\begin{abstract}
We present a time-dependent Floquet method that allows one to use the cycle-averaged Kramers-Henneberger basis for short pulses and arbitrary laser frequencies.
By means of a particular plane-wave expansion we arrive at a time-dependent Schr\"{o}dinger equation that consists of convolutions of momentum and Floquet components. A dedicated numerical treatment of these convolutions, based on Toeplitz matrices and fast Fourier transformations, allows for an efficient time-propagation of large Floquet expansions. 
Three illustrative cases of ionization with different photon energies are analyzed, where the envelope of a short and intense pulse is crucial to the underlying dynamics.

%
%
%

\end{abstract}

\maketitle

\section{Introduction}

Non-perturbative laser-matter interaction provides a rich yet challenging area for theoretical studies. While numerical methods have to deal with large energy bandwidths required to fully account for the dynamics, analytical methods are faced with the challenge of finding an appropriate description of non-perturbative light-matter interaction.

A successful analytical approach to non-perturbative laser-matter interaction is the Kramers-Henneberger (KH) approximation \cite{Henneberger1968a,Gersten1974}. It describes the dynamics in the Kramers-Henneberger reference frame co-moving with the laser-driven electron(s). Traditionally, the time-dependence of the problem is eliminated by using a Hamiltonian, averaged over one optical cycle. The corresponding potential and eigenenergies are often referred to as the KH approach or the KH ``atom." For sufficiently large field strengths and high frequencies, the cycle-averaged Hamiltonian (\CAH) largely determines the properties of the coupled light-matter system, while all higher-order corrections remain small and can be treated perturbatively. 

Since its introduction, the KH approach was thoroughly examined and the properties of \CAH\ are very well known \cite{Gavrila2002, Popov2003}. It was applied to a large variety of problems in atomic, molecular \cite{Pont1988} and solid state physics \cite{Niculescu2008,Lima2008} and in particular, was indispensable in the study of ionization suppression phenomena for atoms in strong and high-frequency fields. Nevertheless, most of the theoretical predictions were not tested experimentally (see \cite{VanDruten1997,DeBoer1993,DeBoer1994} for application for Rydberg state ionization) because high-intensity and high-frequency lasers were not available at that time.

The situation has, however, changed due to the free-electron lasers (FEL) \cite{McNeil2010} that are already able to provide pulses of sufficiently high-frequency and intensity to enable the observation of non-perturbative phenomena. The first experimental studies of Raman processes in the VUV and XUV frequency range, which require coherent multiple photon absorption/emission, have been carried out \cite{Weninger2013}. It can be expected that FEL’s will soon reveal high-frequency non-perturbative phenomena which were proposed theoretically, such as adiabatic stabilization \cite{Gavrila2002}, dynamic interferences \cite{Toyota2008,Toyota2007,Tolstikhin2008,Demekhin2012,Baghery2017}, Rabi oscillations between core-hole states \cite{Demekhin2011}, to name a few.

The KH approach is ideally suited to describe strong-field high-frequency physics to be realized in FEL facilities apart from one crucial aspect: studies so far were mostly limited to continuous-wave laser radiation. Indeed, for a continuous-wave field, a perturbative expansion in Floquet orders can be readily developed. On the other hand, for short (FEL) pulses many Floquet channels have to be included, rendering the time-dependent treatment prohibitively demanding. 

Clearly, the time-dependent aspect is crucial since the short pulses created by FEL sources can lead to additional dynamics driven by the pulse envelope as was recently predicted \cite{Toyota,Simonsen2016}, or be necessary to account for phenomena like impulsive Raman scattering \cite{Miyabe2015}. Hence, in order to apply the KH approach to the dynamics involving intense and short pulses, a formulation different from the ones so far known appears to be necessary. 

Here we propose a numerical approach for short-pulse non-perturbative laser-matter interaction that is based on a time-dependent Floquet formalism in the KH reference frame. It uses \CAHt\ which depends on the instantaneous intensity of the laser pulse and relies on time-propagation using the full Floquet Hamiltonian, which is performed with an efficient Fast-Fourier-Transformation based algorithm.
Combining these two approaches allows us to obtain both a qualitative and quantitative understanding of the light-matter interaction \emph{during} the laser pulse, despite treating short laser pulses produced by FEL facilities non-perturbatively.

In Section \ref{sec:theory} we will present the time-dependent Floquet approach, followed in Section \ref{sec:num} by the introduction of the novel algorithm to solve the Floquet problem in momentum space. The approach is illustrated in Section \ref{sec:results}, where the role of the envelope of a short and intense laser pulse is investigated for the ionization in 1D potential. By varying the laser frequency, while keeping \CAHt\ invariant, three parameter ranges are explored: high, intermediate, and low-frequency regimes. We show in Section \ref{sec:high_freq} that \CAHt\ provides an excellent approximation of the laser-driven dynamics for frequencies higher than the binding energy of the potential. 
For intermediate frequencies close to the ionization threshold, discussed in Section \ref{sec:mid_freq}, the pulse envelope plays a crucial role in determining the channels involved in the ionization.
Finally, the low-frequency regime is discussed in Section \ref{sec:low_freq}; in this case, the photon energy is much smaller than the binding energy of the field-free potential and several hundred Floquet channels are required to fully account for the dynamics. 
We show that the population is rapidly distributed over many excited states of \CAHt\ during the rising part of the laser pulse, which has to be considered if one wants to use the KH approach for low-frequency fields.

\pagebreak
\section{Time-dependent Kramers--Henneberger--Floquet approach in momentum representation} \label{sec:theory}

In this section a time-dependent Floquet approach is derived for efficiently computing laser-driven dynamics with \CAHt\ in the KH reference frame, i.e., using the cycle averaged Hamiltonian that depends on instantaneous laser intensity, and is a generalization of the Envelope Hamiltonian approach introduced in \cite{Toyota}. The formalism allows one to explore the transformation of the wave function from the field-free to the ``field-dressed" picture while still fully accounting for the effects of a short laser pulse.

\subsection{Kramers--Henneberger transformation}

The time-dependent Schr\"{o}dinger equation (TDSE) within the single active electron approximation (in the following, atomic units will be used, unless stated otherwise)
\begin{equation}
\imath \frac{\partial}{\partial t}\Psi(\rv, t) = \Big[-\frac{1}{2} \big( \boldsymbol{\nabla} - \imath \mathbf{A}(t) \big)^2 + V(\rv) \Big] \Psi(\rv, t), \label{eq:tdse}
\end{equation}
where $\mathbf{A}(t)$ relate to the laser electric field $\mathbf{E}(t)$ by $\mathbf{A}(t) = -\int^t\mathrm{d}\tau \, \mathbf{E}(\tau)$, can be transformed into a space-translated frame of reference, the so-called KH frame, by applying the unitary transformation
\begin{align}
\hat{U} = \exp\Big( -\int^t \dd \tau \mathbf{A}(\tau) \cdot \boldsymbol{\nabla}  \Big) \,
\exp\Big( \frac{\imath}{2} \int^t \mathrm{d}\tau \mathbf{A}^2(\tau) \Big).
\end{align}
In the KH reference frame the TDSE acquires the form
\begin{align}
	\imath \frac{\partial}{\partial t} \Psi_{\khsub}(\rv, t) = \Big[-\frac{\boldsymbol{\nabla}^2}{2} + V\big(\rv + \av(t)\big) \Big] \Psi_{\khsub}(\rv, t), \label{eq:tdse_kh}
\end{align}
where the coupling with the laser field is reduced to the time-dependent shift $\av(t)$ of the binding potential $V\big(\rv + \av(t)\big)$. For simplicity we assume this shift to be of the form
	\begin{align}
	\av(t) = -\int^t \mathbf{A}(\tau) \, \mathrm{d}\tau = \av_0(t) \, \cos(\omega \, t + \phase)
	\label{eq:excursion}
	\end{align}
corresponding to the classical trajectory of a charged particle in a laser field. 


\begin{figure}
\centering
\includegraphics{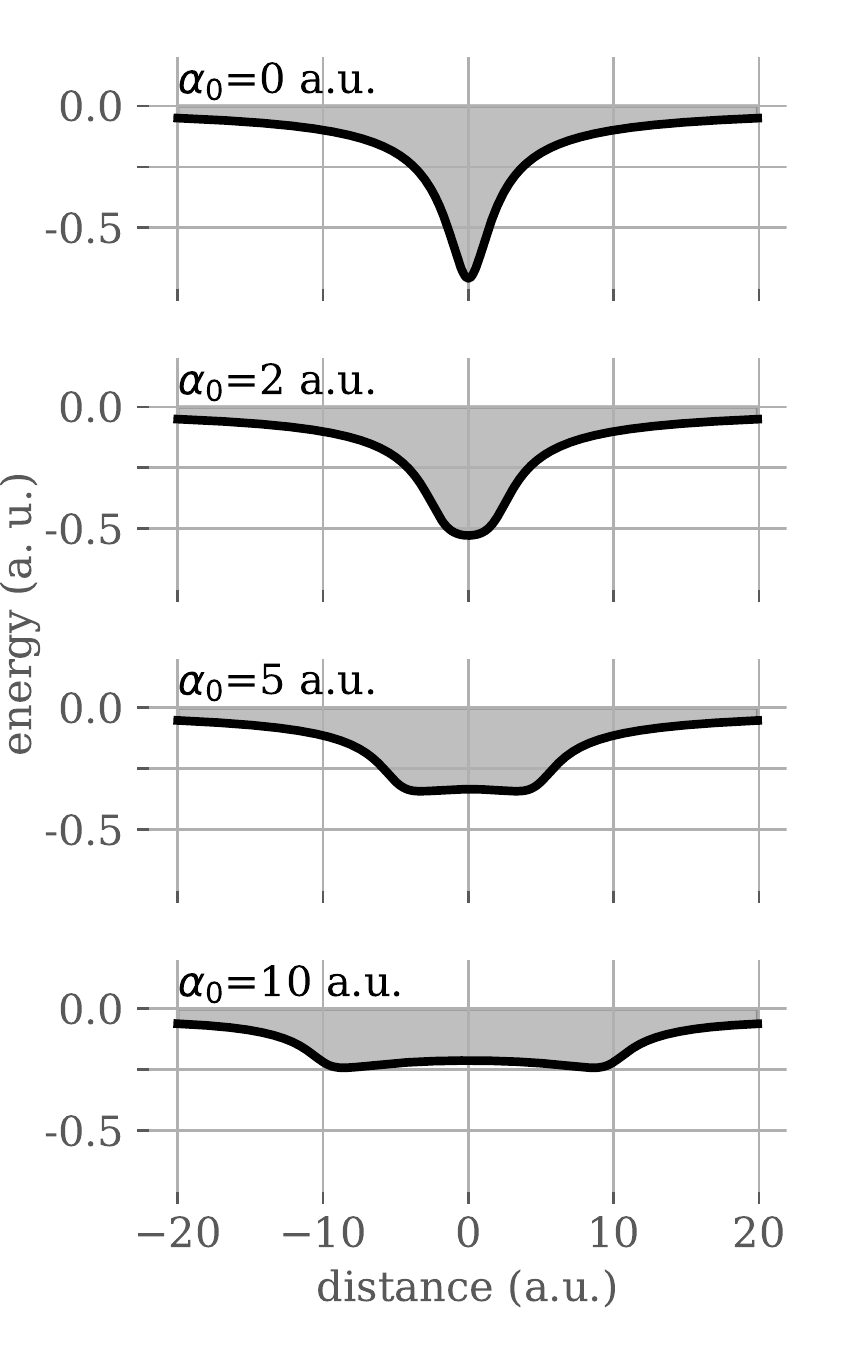}
\caption{Cycle-averaged soft-core atomic potential (see Eq.~(\ref{eq:V0}) and Eq.~(\ref{eq:V_soft}) ) for different excursion lengths $\as_0$. Shaded area indicates binding part of the potential.}
\label{fig:KH_potentials}
\end{figure}

The KH transformation describes the laser--atom interaction in a frame of reference, where the electron can be considered to be ``stationary," while the binding potential of the ``atom" is time-dependent. In other words, the electron ``sees the nucleus oscillating back and forth". The oscillating potential $V\big(\rv + \av(t)\big)$ can be integrated over a single cycle $T_\omega$ of the oscillation $\av(t)$ to obtain the averaged potential
\begin{equation}
V_{0}(\rv, \av_0) = \frac{1}{T_\omega}\int_0^{T_\omega} \, \dd t \, V\big(\rv + \av(t)\big), \label{eq:V0}
\end{equation}
which is also called the ``KH potential" and has been used to describe the properties of atoms in strong and high-frequency laser fields \cite{Pont1988,Pont1990}. The average potential strongly depends on the electron excursion length $\av_0$, as illustrated in \fig \ref{fig:KH_potentials}, and, for sufficiently large excursion lengths $\av_0$, transforms from a single-well to a dichotomous double-well shape.
In this work, we are going to use the cycle-averaged potential that adjusts to time-variation of $\av_0(t)$ to describe the atom-laser interaction with pulsed laser fields.

\subsection{Time-dependent Floquet approach for short laser pulses}

The time-independent KH potential and its properties are analyzed in great detail in the literature using a variety of methods \cite{Gavrila1984,Reed1990,Pont1990,Su1990,Dorr1991,Yao1992,Scrinzi1993,Atabek1997,Smirnova2000,Morales2011}.
In practice, however, one needs to deal with finite and often short pulses and considering a static KH potential is not sufficient.
Here, we consider a cycle-averaged potential that adjusts to the laser pulse envelope, while still providing an exact description of the dynamics. At first glance it looks cumbersome to perform for each instance of time a full cycle average. However, when switching simultaneously to momentum space one arrives at a compact and distinct form of the TDSE (cf. Eq.~(\ref{eq:tdse_tt}) below), as we will show briefly here and in detail in Appendix \ref{sec:app_expansions}.

The potential in the KH reference frame can be written in a plane-wave expansion
\begin{align}
V(\rv,t) & = \int\dd^{3}r'\!\int\dd t'\,V(\rv',t')\,\delta(\rv'{-}\rv)\,\delta(t'{-}t)
\notag\\
&= \KNORM \WNORM \sum_{m} \intK \;
\Big[\int \dd^{3}r'\!\int_{0}^{T_\omega}\dd t'\,V(\rv',t')\e{+\imath m\omega t'}\e{-\imath \kv \cdot \rv'}\Big]
\e{-\imath m\omega t}\e{+\imath \kv \cdot \rv}
\label{eq:expa}
\end{align}
with integer $m$, anticipating that the potential oscillates with frequency $\omega$.

In order to efficiently treat short pulses we will not apply the standard expansion, but rather extract the envelope of the laser pulse.
This is done by splitting the electron displacement $\av(t)$ into the non-periodic envelope $\av_0(t)$, described by the time variable $t$ and the periodic oscillation $\cos(\omega t'+\phase)$, described by time $t'$. Thereby, the KH potential becomes a ``two-time potential''
\begin{equation}
\label{eq:ttpot}
\Vb(\rv, t, t') = V\big(\rv + \av_{0}(t)\cos(\omega t'{+}\phase)\big),
\end{equation}
which can be used straightaway in expansion \eqref{eq:expa} to give
\begin{subequations}\label{eq:ttexpa}
\begin{align}
V(\rv,t) &= \sum_{m} \intK \; \Vmk \,
\e{-\imath m\omega t}\e{+\imath \kv{\cdot}\rv}
\intertext{with}
\Vmk &\equiv 
\KNORM \WNORM
\int \dd^{3}r'\!\int_{0}^{T_\omega}\dd t'\,\Vb(\rv', t, t')\e{+\imath m\omega t'}\e{-\imath \kv{\cdot}\rv'}.
\end{align}
\end{subequations}
Thereby, the components $\Vmk$ depend on time through the pulse envelope.
Needless to say that expansion \eqref{eq:ttexpa} is exact.
There are other approaches that adopt two times \cite{Peskin1993,Telnov1995,Chu2010a,Halasz2012}. Here the two times are used to straightforwardly derive expansion (\ref{eq:ttexpa}), which turn out to be very convenient for short pulses.

Now it is essential that in the KH reference frame, by means of a translation in space and the Jacobi-Anger expansion, the components $\Vmk$ can be rewritten as products (see Appendix~\ref{sec:app_expansions} and Ref.~\cite{Yao1992} for the derivation)
\begin{subequations}\label{eq:VJexp}
\begin{align}
\Vmk &=\Vk \,\imath^{|m|} J_{|m|} \big(\kv\,{\cdot}\,\av_{0}(t)\big)\e{-\imath m\phi}
\intertext{with the momentum components of the field-free potential}
\Vk &\equiv \KNORM \int \dd^{3}r\,V(\rv)\,\e{-\imath\kv{\cdot}\rv}
\end{align}
\end{subequations}
and $J_{m}$ denoting the ordinary Bessel functions of the 1st kind.

Having rewritten the potential as a sum of products we use a similar expansion in terms of Fourier modes and plane-waves for the wave function
\begin{equation}
\Psi_{\khsub}(\rv, t) = \sum_{m} \intK \; \Pmk \, \mathrm{e}^{-\imath m \omega t} \, \mathrm{e}^{\imath \kv \cdot \rv}. \label{eq:ansatz_wf_fourier}
\end{equation}
Just like in the Eq.~(\ref{eq:ttexpa}), this ansatz does not imply any restriction on the total wave function. The expansion coefficients $\Pmk$ are determined by inserting Eqs.~(\ref{eq:ttexpa},\ref{eq:VJexp}) and Eq.~(\ref{eq:ansatz_wf_fourier}) into the TDSE. This allows one to derive an equation for the $\kv$-th plane-wave and $m$-th Fourier components of the wave function (see Appendix \ref{sec:app_expansions} for details)
\begin{multline}
\imath \frac{\partial}{\partial t}\Pmk = \Big[ \frac{\kv^2}{2} - m \omega \Big] \Pmk +\\
+ \sum_{m'} \intK' \; \Vkp \, \Pmpkp \;  
\imath^{|m-m'|} \, J_{|m-m'|}\big( (\kv-\kv') \cdot \av_0(t) \big) \; \mathrm{e}^{-\imath (m-m')\phase}.\label{eq:tdse_tt}
\end{multline}
Eq.~(\ref{eq:tdse_tt}) is the main equation used in this work and provides an \emph{exact} description of the laser driven dynamics in the KH reference frame. 

The accuracy of its numerical implementation is limited only by the basis and propagation routines, see Sec.~\ref{sec:accuracy} for more extended discussion.
The momentum representation used here is particularly suited to describe the dynamics in the KH reference frame as it reduces the TDSE to a convenient form that allows one to use efficient numerical propagation methods, as described in Section \ref{sec:num}.

\subsubsection{Physical interpretation of the wave function in the Fourier basis}

The physical significance of the index $m$ becomes apparent, if we consider an isolated Fourier subspace $m$, i.e., ignore the coupling between the wave function coefficients $\Pmk$ with different $m$. In such a case, the only remaining potential coupling terms in Eq.~(\ref{eq:tdse_tt}) are
\begin{equation}
\Vkpo = \Vkp \, J_{0}\big( (\kv-\kv') \cdot \av_0(t) \big),
\end{equation}
which is just the momentum-representation of the cycle-averaged potential
\begin{equation}
\intK \; \Vko \, \e{\imath \kv \cdot \rv} =  \frac{1}{T_\omega}\int_0^{T_\omega} \, \dd t' \, V\big(\rv + \av(t,t')\big). 
\end{equation}
Therefore, considering a single Fourier subspace in isolation is similar to the original KH approach \cite{Henneberger1968a}, where only the cycle-averaged potential is considered.

The components $\Vmk$ with $|m|>0$ couple different Fourier subspaces and lead to transitions between the states of the cycle-averaged Hamiltonian \CAHt. Therefore, the index $m$ can be interpreted as the number of absorbed/emitted photons. For example, population initially created in the $m=0$ subspace and ending up in the $m$-th subspace after the pulse represents $m$-photon absorption, therefore, when its physical meaning will be important the $m$-th Fourier subspace will be referred to as Floquet channel.
In this work, enough Floquet channels are included to achieve numerical convergence. Therefore fields of arbitrary frequency can be considered.

\section{Numerical implementation} \label{sec:num}

\editt{
To numerically solve Eq.~(\ref{eq:tdse_tt})  we first rewrite it for a discrete momentum $\kv$ grid, yielding
\begin{multline}
\imath \frac{\partial}{\partial t}\Pdmk = \Big[ \frac{\kv^2}{2} - m \omega \Big] \Pdmk +\\
+ \sum_{\kv' m'} \Vdkp \, \Pdmkp \;  
\imath^{|m-m'|} \, J_{|m-m'|}\big( (\kv-\kv') \cdot \av_0(t) \big) \; \mathrm{e}^{-\imath (m-m')\phase}, \label{eq:tdse_tt_disc}
\end{multline}
where for D dimensions the field-free potential and the wave function are renormalized according to $\Vdk {=} \widetilde{V}(\kv) \, (\dk)^D$ and $\Pdmk=\Pmk (\dk)^{D/2}$
implying a box discretization with a box of size $L^D=(2\pi/\dk)^D$.
}
\edit{
The right-hand side of Eq.~(\ref{eq:tdse_tt_disc}) can be split into two parts. The first part, which using matrix notation is defined by
\begin{equation}
[ \Tmat \cdot \Pvec ]_{\kv m} \equiv \Big[ \frac{\kv^2}{2} - m \omega \Big] \Pdmk,
\end{equation}
is diagonal and can be easily computed numerically. The computation of the sum
\begin{equation}
[ \Vmat \cdot \Pvec ]_{\kv m} \equiv \sum_{\kv' m'} \Vdkp \Pdmkp \times  
\imath^{|m-m'|} \, J_{|m-m'|}\big( (\kv-\kv') \cdot \av_0(t) \big) \times \mathrm{e}^{-\imath (m-m')\phase} \label{eq:potential_operator}
\end{equation}
requires the main numerical effort as it is associated with the non-diagonal elements of $\Vmat$.
In the field-free case ($\av_0=0$), the part (\ref{eq:potential_operator}) describes a convolution between momentum components $\kv$ of the wave function  and the potential. If the laser field is present ($\av_0 \neq 0$), additional terms proportional to $ J_{|m-m'|}\big( (\kv-\kv') \cdot \av_0(t) \big)$ enter the sum (\ref{eq:potential_operator}). They couple different Floquet channels $m$ and also modify the coupling between momentum components $\kv$. Nevertheless, the convolution form of the matrix $\Vmat$ in (\ref{eq:potential_operator}) is preserved, since the couplings depend only on the differences $\kv-\kv'$ and $m-m'$. Note, that $\Vmat$ and $\Pvec$ depend on time, which will be kept implicit for the brevity of notation. 
}

\edit{
The convolution form of the matrix $\Vmat$ allows one to apply the convolution theorem and to replace the convolution between potential and wave function,
described by (\ref{eq:potential_operator}), by their product in the Fourier domain. This greatly increases the speed of computation, in particular if a fast Fourier transformation (FFT) algorithm is used to convert to and from the Fourier domain.
}

\edit{
The convolution theorem strictly holds only for infinite or periodic vectors, which implies an expansion in $\kv$ and $m$ to infinite order. In practical numerical calculations, the necessity to use a finite size basis will normally violate the conditions for validity of the convolution theorem, consequently causing numerical errors.
Therefore,
we use an alternative approach that is based on the theory of Toeplitz matrices \cite{doi:10.1137/1.9780898718850}. It takes advantage of the convolution form of the matrix $\Vmat$ and allows one to use the FFT algorithm to accelerate the calculations. However, unlike the direct application of convolution theorem, the method based on the Toeplitz matrix theory is exact for vectors of finite size. 
This approach is particularly useful to study Floquet systems, as it allows one to truncate the basis to only a few Floquet channels. 
}

\subsubsection{Description of the algorithm}

For a single Floquet channel, e.g., $m=m'$, the elements along the diagonal of the matrix $\Vmat$ in Eq.~(\ref{eq:potential_operator}) are equal, which follows directly from the momentum representation. Such a matrix is called a Toeplitz matrix and its properties are well known in the literature, see, e.g., \cite{doi:10.1137/1.9780898718850}. It can be fully described by a single row and column only. Furthermore, a product of a finite Toeplitz matrix with any vector can be performed exactly using the FFT algorithm.

The algorithm to multiply a Toeplitz matrix $\Vmat$ with a vector $\Pvec$ is as follows (see Appendix \ref{sec:app_T} for a more detailed description): 
\begin{enumerate}
	\item A circulant vector $\Cvec$ is formed from the first column and the first row of the matrix $\Vmat$.$\bf$
	\item Zeros are appended to the vector $\Pvec$ to match the length of $\Cvec$.
	\item A Fourier transformation of both the circulant vector $\Cvec$ and the extended coefficient vector $\Pvec$ is performed.
	\item The two transformed vectors are multiplied and an inverse Fourier transformation is applied to the product.
\end{enumerate}
The first half of the final vector now stores the matrix--vector multiplication result, while the second half is discarded.

If the couplings between different Floquet channels are considered, i.e., $m-m' \neq 0$, then the matrix $\Vmat$ is of block form with all equal blocks on the same diagonal. Additionally, each block is of Toeplitz form. Such a matrix is called a Block Toeplitz matrix with Toeplitz Blocks (BTTB). The product of a BTTB matrix and any vector can be performed using a two-dimensional Fourier transformation algorithm in a similar way as a Toeplitz matrix--vector product, see Appendix \ref{sec:app_BTTB} for a detailed description. The approach can be further extended to an arbitrary number of dimensions. 

The algorithm to calculate the Toeplitz matrix-vector product can be considered as generalization of the well-known split operator technique (transformation to Fourier domain, multiplication and inverse transformation), see, e.g., \cite{Kosloff1988}, which is widely used to solve the TDSE. On the other hand, the algorithm presented here cannot be used to directly evaluate the product of a vector with a function of Toeplitz matrix, e.g., $ \exp(-\imath \Vmat \Delta t) \Pvec$. \hide{Nevertheless, our implementation reduces the number of Fourier and plane-wave components required to achieve high numerical accuracy. Therefore, for finite systems or truncated bases the Toeplitz matrix-vector multiplication algorithm outperforms the traditional split-operator technique.}
\editt{Nevertheless, the Toeplitz matrix-vector multiplication algorithm allows us to reduce the number of Fourier and plane-wave components required to achieve high numerical accuracy and allows it to outperform the traditional split-operator technique.}

\subsubsection{Time propagation}

Many different numerical methods to solve Eq.~(\ref{eq:tdse_tt_disc}) could be used, for example explicit Runge-Kutta or Arnoldi-Krylov algorithms. 
However, to take advantage of the BTTB symmetry of the potential matrix $\Vmat$, the matrix-vector multiplications involving $\Vmat$ must be implemented using efficient FFT routines with the method described above.
In this work the Taylor expansion propagator is used. This method relies on the expansion of the propagator over a discrete time-step $\Delta t$ in a Taylor series up to the desired order, so that the wave function expansion coefficients can be computed as
\begin{align}
\Pvec(t+\Delta t) = \exp\big[ -\imath ( \Tmat + \Vmat )\Delta t \big] \Pvec(t) = \nonumber\\
[ \mathbf{1} - 
\imath ( \Tmat + \Vmat )\Delta t 
-  \frac{1}{2}( \Tmat + \Vmat )^2 \Delta t^2 + \ldots] \Pvec(t),\label{eq:Taylor_expansion_propagator}
\end{align}
where each term in the expansion is evaluated iteratively. Hence, the numerical problem reduces to the evaluation of products $\Tmat\cdot\Pvec$, where $\mathbf{T}$ is diagonal, and $\Vmat\cdot\Pvec$, which is evaluated using the Toeplitz matrix--vector multiplication algorithm presented above. 
The accuracy can be controlled by choosing  the order of expansion at each time-step.
Although the propagator is not norm-conserving, if enough expansion orders are included norm conservation up to a desired numerical accuracy can be easily achieved.
In this work, the expansion was truncated once the norm of the corrections to the wave function coefficients dropped below $10^{-16}$.
The Taylor expansion propagator thus provides an accurate and reliable method to obtain a numerical solution to Eq.~(\ref{eq:tdse_tt}). Importantly, combined with the FFT algorithm for matrix-vector multiplication operations, large Fourier expansion orders $m$ can be treated explicitly.
More sophisticated propagation methods that also rely on matrix-vector products like Arnoldi-Krylov-propagators may be easily implemented.

\subsection{Accuracy} \label{sec:accuracy}

\begin{figure}
\centering
\includegraphics{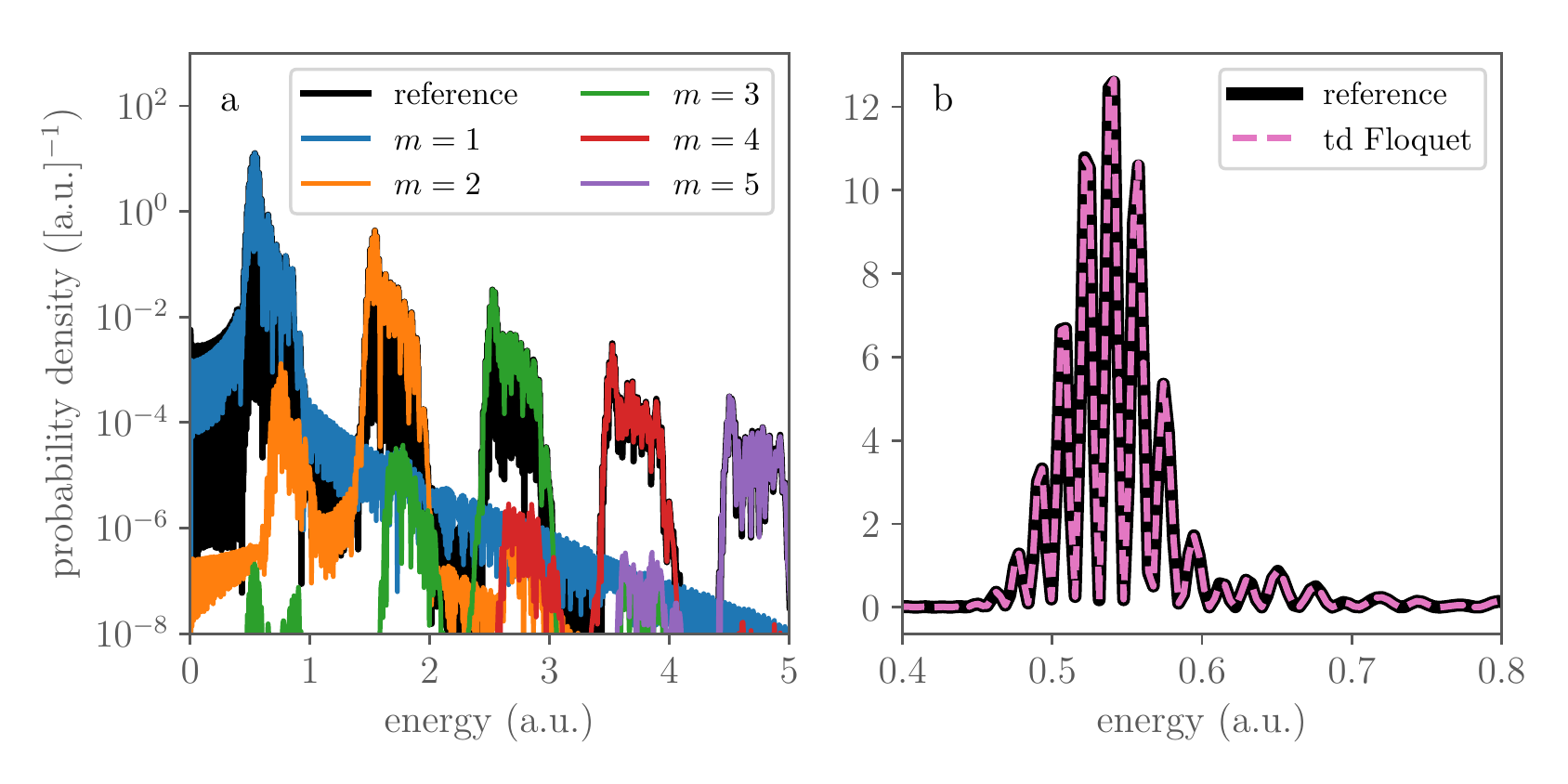}
\caption{Energy resolved photoionization spectra after the pulse obtained by solving the TDSE in velocity gauge (black lines) compared with the spectra obtained using the time-dependent Floquet approach (a) for each Floquet channel $m$; (b) combined spectra from all Floquet channels.}
\label{fig:spectra_benchmark}
\end{figure}

The accuracy of the time-dependent Floquet approach developed in this work is verified by comparing the wave function obtained by directly solving the TDSE in velocity gauge in Eq.~(\ref{eq:tdse}) with the solution of the TDSE defined in Eq.~(\ref{eq:tdse_tt_disc}). In both cases, identical plane-wave basis and propagator routines of the TDSE were used. 

For all laser pulse parameters that were used in this work, the wave functions obtained from the time-dependent Floquet approach and by directly solving the TDSE in velocity gauge were found to match up to numerical accuracy, if sufficiently many Floquet channels $m$ were considered.
The accuracy of the time-propagation procedure is determined by the time-step and Taylor expansion order. The required number of Fourier components $m_{max}$ can be determined from the plane-wave basis set by requiring that $m_{max} \, \omega > |\kv|^2_{max}/2$, where $|\kv|_{max}$ is maximum momenta described by the plane-wave basis. Note, however, that in the Floquet formulation of TDSE in Eq.~(\ref{eq:tdse_tt}) both positive and negative Floquet channels have to be considered, i.e., $-m_{max} < m < m_{max}$.

An illustrative example is provided in Fig.~\ref{fig:spectra_benchmark} for ionization from a soft-core potential, which is defined in Sec.~\ref{sec:model}, with $\omega=1$ a.u. photon energy, $I=2.4 \times 10^{18} \, \text{W/cm}^2$ intensity and 5 fs full-width at half-maximum duration pulse. The spectra under similar laser pulse parameters were extensively investigated in previous works \cite{Toyota2008,Toyota2007,Tolstikhin2008,Demekhin2012,Baghery2017} and the calculation is further discussed in Sec.~\ref{sec:high_freq}, therefore here we only note that each Floquet channel provides the $m$-photon absorption channel, see Fig.~\ref{fig:spectra_benchmark}a. The final spectra, obtained by summation over all Floquet channels $m$, is indistinguishable from the spectra obtained by the direct solution of the TDSE in velocity gauge, see  Fig.~\ref{fig:spectra_benchmark}b. 

The approach was tested to be accurate for photon energies ranging from 0.05 to 1 a.u. Furthermore, it was accurate for pulses down to single cycle duration for both low and high frequencies. 
Therefore, the time-dependent Floquet formalism is capable of fully describing the dynamics driven by intense and short laser pulses using the cycle integrated Hamiltonian for arbitrary laser parameters.

The accuracy of the numerical procedure is further dictated by the quality of the plane-wave basis set. In all the calculations presented in this work, a converged basis set in terms of maximum momenta $|\kv|_{max}$ and spacing between momenta components $\dk$ is used. 

Finally, note that atomic potentials with a long-range tail lead to a singularity at the origin in the momentum representation, i.e., $\Vt(\kv=0) \rightarrow -\infty$. This singularity could be treated by, for example, a Land\`{e} subtraction  procedure \cite{Norbury1994,Shvetsov-Shilovski2014}. Alternatively, the potential can be considered in a ``finite box", as in \cite{Jiang2008}. Finally, since adding a delta function to a momentum representation of a potential leads only to a trivial energy shift of the spectra, the singularity can be removed by introducing a new potential $\Vdkp + \xi \delta_{\kv-\kv'}$ into Eq.~(\ref{eq:tdse_tt_disc}) with $\xi \rightarrow \infty$, where $\delta_{\kv-\kv'}$ is the Kronecker delta function. Numerically this corresponds to setting all elements  $ \Vidx{\kv-\kv'=0}$ to zero.

\subsection{Performance}

\begin{figure}
	\centering
	\includegraphics{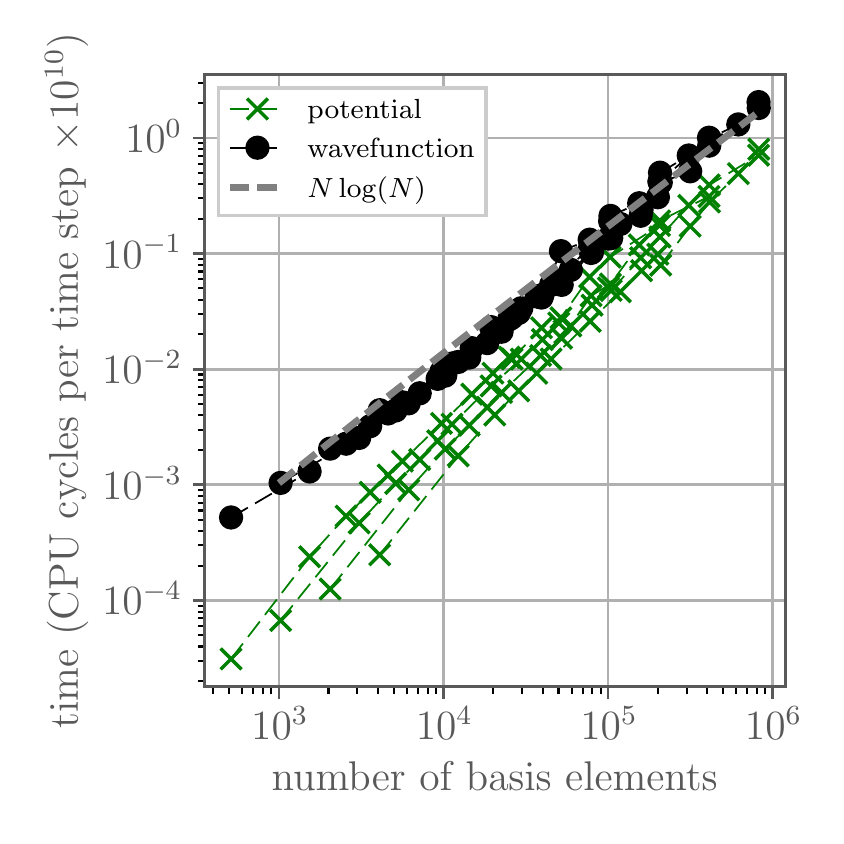}
	\caption{Numerical effort as a function of the total number of basis elements, evaluated in terms of processor cycles spent per single time-step for calculations with different basis sizes and pulse durations. Black dots indicate the effort required to compute a single time-step; green crosses -- time effort required to update the Hamiltonian. Gray dashed line indicates the scaling $N\log(N)$.}
	\label{fig:scaling_element_number}
\end{figure}

\begin{figure}
\centering
\includegraphics{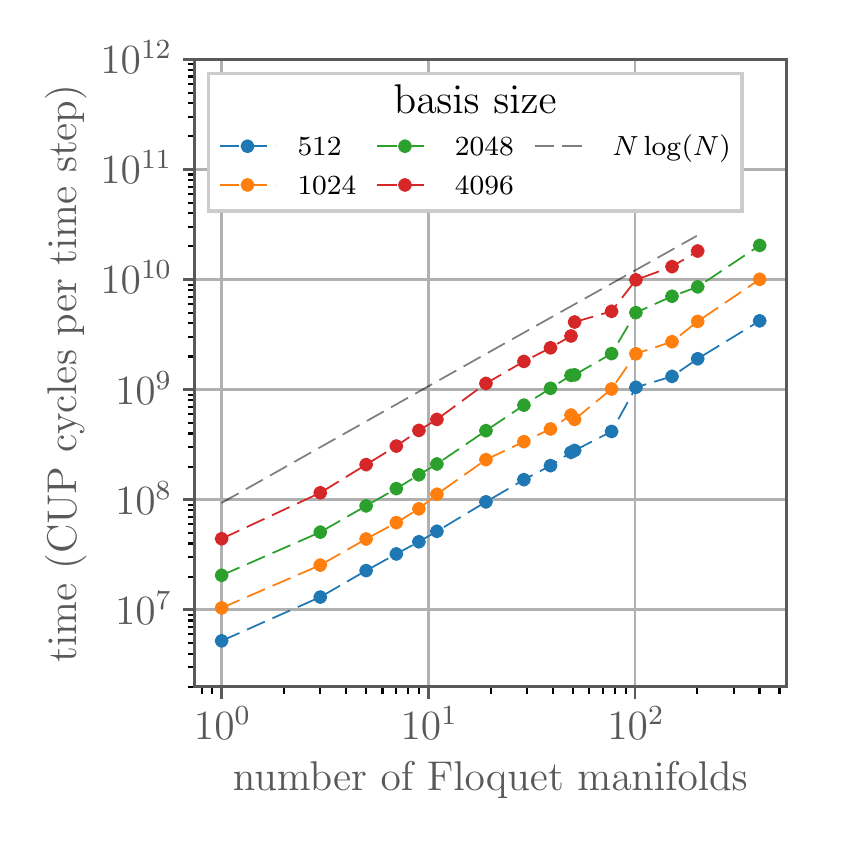}
\caption{Numerical effort as a function of the number of Fourier components for different plane-wave basis sizes, evaluated in terms of processor cycles spent per single time-step; gray dashed line indicates the scaling $N\log(N)$.}
\label{fig:scaling_basis_size}
\end{figure}

The Toeplitz matrix approach described above allows to efficiently solve the time-dependent Floquet formulation of TDSE in Eq.~(\ref{eq:tdse_tt_disc}) using the FFT based matrix-vector multiplication. The use of FFT algorithm allows to achieve scaling proportional to $N\log(N)$ with respect to the size of the basis $N=N_K \times N_F$, where $N_K$ is the number of plane-waves and $N_F$ is the number of Fourier components. 
This scaling is illustrated as a function of the total number of basis elements $N$ in Fig.~\ref{fig:scaling_element_number} for different calculations with laser pulse parameters used in this work. The size of the plane-waves basis is varied between 512 and 4096 with the maximum momenta kept fixed. The number of Fourier components is varied between 1 and 401. The numerical effort is measured in terms of processor cycles spent solving the TDSE. The number of cycles is then divided by the total number of time-steps used, so that calculations using different pulse lengths could be directly compared. In addition, the expansion order of the Taylor propagator in Eq.~(\ref{eq:Taylor_expansion_propagator}) is kept fixed at 8. In an adaptive expansion scheme, the expansion order mainly depends on $|\kv|_{max} \times \Delta t$.

The main effort required to solve the TDSE stems from updating the wave function at each time-step using the Taylor expansion method, which is illustrated by the black dots in Fig.~\ref{fig:scaling_element_number}. It scales proportionally to $N \, \log(N)$ as expected. The numerical effort required for different sizes of the plane-wave basis is shown in Fig.~\ref{fig:scaling_basis_size}. Again, the effort scales proportionally to $~N_F \, \log(N_F)$ with the number of Fourier components.
Additional numerical effort is required to update the elements of potential energy operator at every time step, since they depend on the laser field. This effort is illustrated by green crosses in Fig.~\ref{fig:scaling_element_number}. It can take up to 40\% of the total effort. However, it scales linearly with the total number of basis elements since only $N_K \times N_F$ elements are stored in memory.

The time-dependent Floquet method presented here cannot compete in efficiency with conventional approaches to solve TDSE that do not use Floquet expansion. The numerical effort required for the latter would be comparable to using just a single Floquet channel, see Fig.~\ref{fig:scaling_basis_size}. However, the time-dependent Floquet method does not aim to compete with the established approaches in terms of speed or accuracy, but rather to provide an efficient way to tackle large-scale Floquet problems. Therefore, the strength of the current approach is its ability to provide insight into the dynamics during the laser pulse, which is possible due to Floquet-like approach only.

\subsubsection{Extension to more spatial dimensions}

Although this work is limited to one-dimensional potentials, the generalization to more dimensions $D$ for a  plane-wave basis in Cartesian coordinates is straightforward. However, such an approach does not take advantage of the symmetry of the potential and therefore in general requires a large number of basis elements to be included into the Hamiltonian, which scales as $N_K^D \times N_F$. The corresponding increase of numerical effort can be extrapolated from Fig.~\ref{fig:scaling_element_number} and  Fig.~\ref{fig:scaling_basis_size}. 

A direct extension of the method to, e.g., a spherical coordinate system is not straightforward. The advantage of the plane-wave basis in Cartesian coordinates is the separation of any arbitrary potential in the KH reference frame into time-independent and time-dependent parts, as can be seen from Eq.~(\ref{eq:VJexp}), which allows us to calculate the coupling between the plane-wave components at each time-step efficiently. 
We did not find such a simple form for the expansion of the KH potential into spherical harmonics for linearly polarized fields. 

A possible alternative approach to describe atoms in linearly polarized laser field beyond a single dimension is to use cylindrical coordinate system (see, e.g., \cite{Pont1990a}). Since the KH potential is symmetric around the laser polarization axis, a plane-wave expansion can be applied along this direction. 
The KH approach can also be formulated for a circularly polarized field, see, e.g., \cite{Pont1990a}.
Finally, multi-pole expansion of the KH potential can be used, which allows for an efficient description using conventional quantum chemistry methods \cite{Atabek1997}.

\pagebreak
\section{Dynamics driven by short laser pulses using the Kramers--Henneberger--Floquet representation} \label{sec:results}

\subsection{Model system} \label{sec:model}

\begin{figure*}
	\centering
	\includegraphics[]{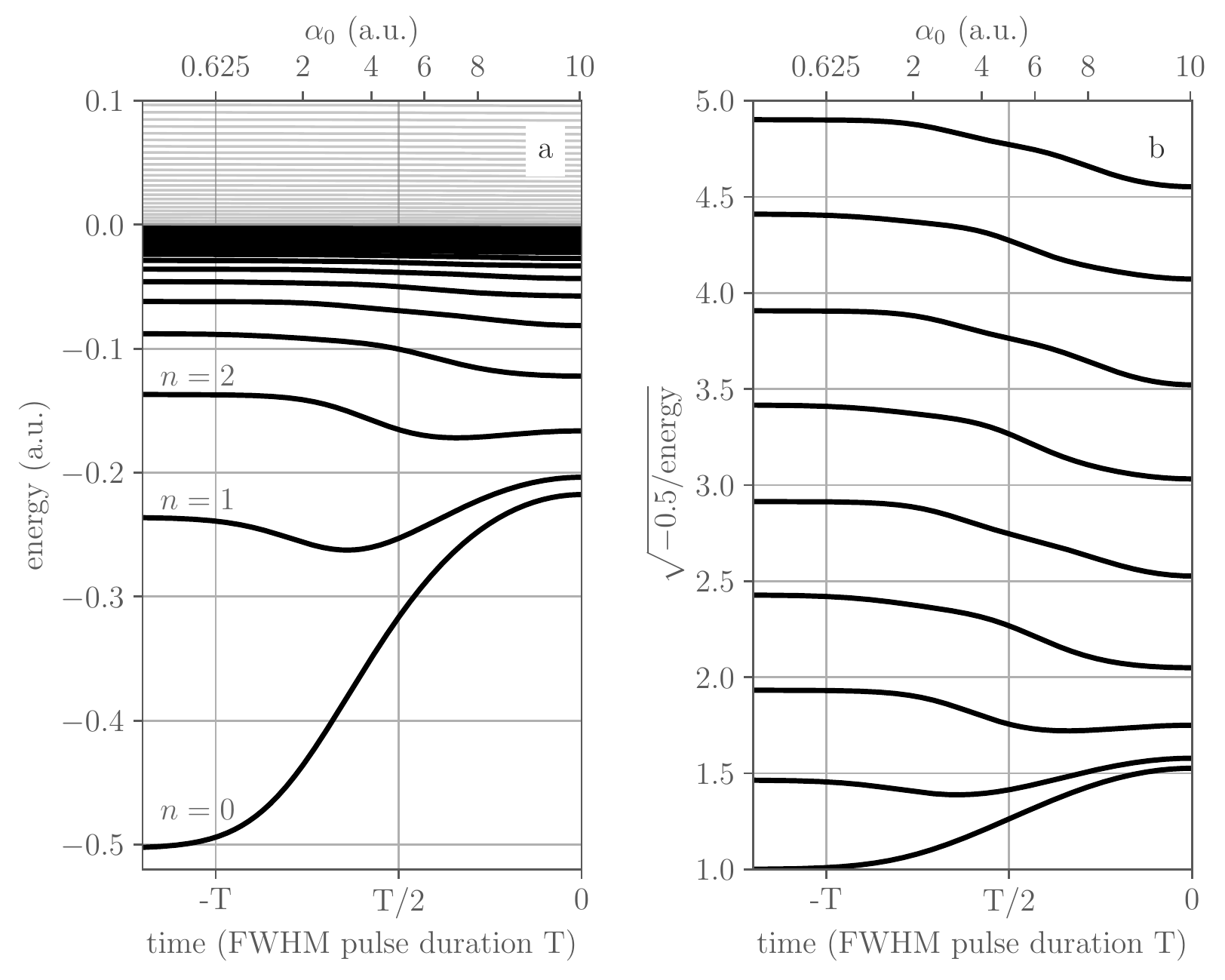}
	\caption{
	\editt{
	Eigenenergies (a) and effective quantum numbers (b) of the cycle-averaged soft-core potential (see Eqs.~(\ref{eq:V0}) and (\ref{eq:V_soft})) as a function of time along the pulse envelope (bottom axis) and classical excursion length $\as_0$ (top axis) for a maximal excursion length of $\as_0=10$ a.u. and a pulse defined in Eq.~(\ref{eq:pulse}).
	}
	}
	\label{fig:Ladder}
\end{figure*}


The time-dependent Floquet approach is illustrated using the example of a 1D model atom, described by a soft-core potential
\begin{equation}
V(x) = -\frac{1}{\sqrt{x^2+x_0^2}}, \label{eq:V_soft}
\end{equation}
which has been widely used to study the dynamics of atoms in high intensity laser fields both analytically and numerically \cite{Su1990,Su1991}. In this work the softening parameter is chosen to be $x_0^2=2$, which leads to a binding energy equal to that of a hydrogen atom $I_p$=0.5 a.u. 

The laser pulse with a peak electric field $F_0$ is defined in terms of the classical electron trajectory introduced in Eq.~(\ref{eq:excursion})  with a Gaussian envelope function \cite{Toyota}
\begin{subequations}\label{eq:pulse}
\begin{equation}
\alpha(t) = \frac{F_0}{\omega^2} \, P\left( T\omega \right)  \, \exp\left( -4 \ln2 \left(t/T\right)^2 \right) \, \cos(\omega \, t + \phase),
\end{equation}
\begin{equation}
P\left( T\omega \right) =  \frac{1}{1+8\ln2/(T\omega)^2}.
\end{equation}
\end{subequations}
An envelope of $T=5$ fs full-width at half-maximum (FWHM) is used, unless specified otherwise. For all except few cycle pulses $ P\left( T\omega \right) \sim 1$ holds.

Furthermore, laser intensity and frequency are chosen such that the maximum classical excursion length $\as_0$ is equal to 
\begin{equation}
\alpha_0 = \frac{F_0}{\omega^2} = 10 \, \text{a.u.}
\end{equation}
for all laser frequencies investigated. Since the KH potential depends only on the classical excursion length $\as_0$  the eigenenergies of the cycle-averaged potential will have identical time-dependence. Nevertheless, the dynamics will still depend on the frequency via the spacing between Floquet channels.  Therefore, the choice of a constant maximal $\as_0$ will allow one to clearly separate the role of the cycle-averaged potential from the role of the laser frequency.


The typical eigenenergy spectra of the 1D cycle-averaged potential $V_0(x, \as_0)$ are depicted in \fig \ref{fig:Ladder}a as a function of time during the pulse. They are obtained by diagonalizing the Hamiltonian in a single Floquet channel. 
The eigenenergies of \CAHt\ strongly depend on the instantaneous intensity of the laser pulse due to the widening and the formation of the dichotomy of the cycle-averaged potential in \fig \ref{fig:KH_potentials} for increasing electron excursion $\as_0$. 

\editt{
In \fig \ref{fig:Ladder}b the effective quantum numbers $n^*=\sqrt{-0.5/E_n}$ are plotted for bound states of \CAHt\ as a function of time along the laser pulse. Bound states of a hydrogenic potential would lead to an infinite series of equally spaced $n^*$, which is the case at the initial time in \fig \ref{fig:Ladder}b. Deviation from a pure hydrogenic potential lead to an uneven spacing of $n^*$, referred to as quantum defect. In the case of the cycle-averaged potential, the quantum defect is a result of the dichotomy of the potential and is clearly visible for the lowest eigenstates. On the other hand, although the energy of the higher $(n>3)$ eigenstates is lowered due to the widening of the cycle-averaged potential, $n^*$ stay approximately equidistant, indicating that these states are determined by the long-range Coulomb tail of the cycle-averaged potential and are not strongly influenced by its dichotomy.}

All the calculations presented further in this work were done using 2048 plane-wave basis states with momenta equidistantly spaced by $\dk = 2\pi / 2000$ a.u. A time-step of $\Delta t = 0.1$ a.u. was used for the propagation, which we found sufficient to obtain converged ionization probabilities. The ground state was obtained by diagonalizing the Hamiltonian defined by Eq.~(\ref{eq:tdse_tt_disc}) with $\as_0=0$ for a single $m=0$ Floquet channel. Finally, the carrier-envelope phase was set to $\phase=0$ in all calculations.

\subsection{High-frequency} \label{sec:high_freq}

\begin{figure*}
	\centering
	\includegraphics{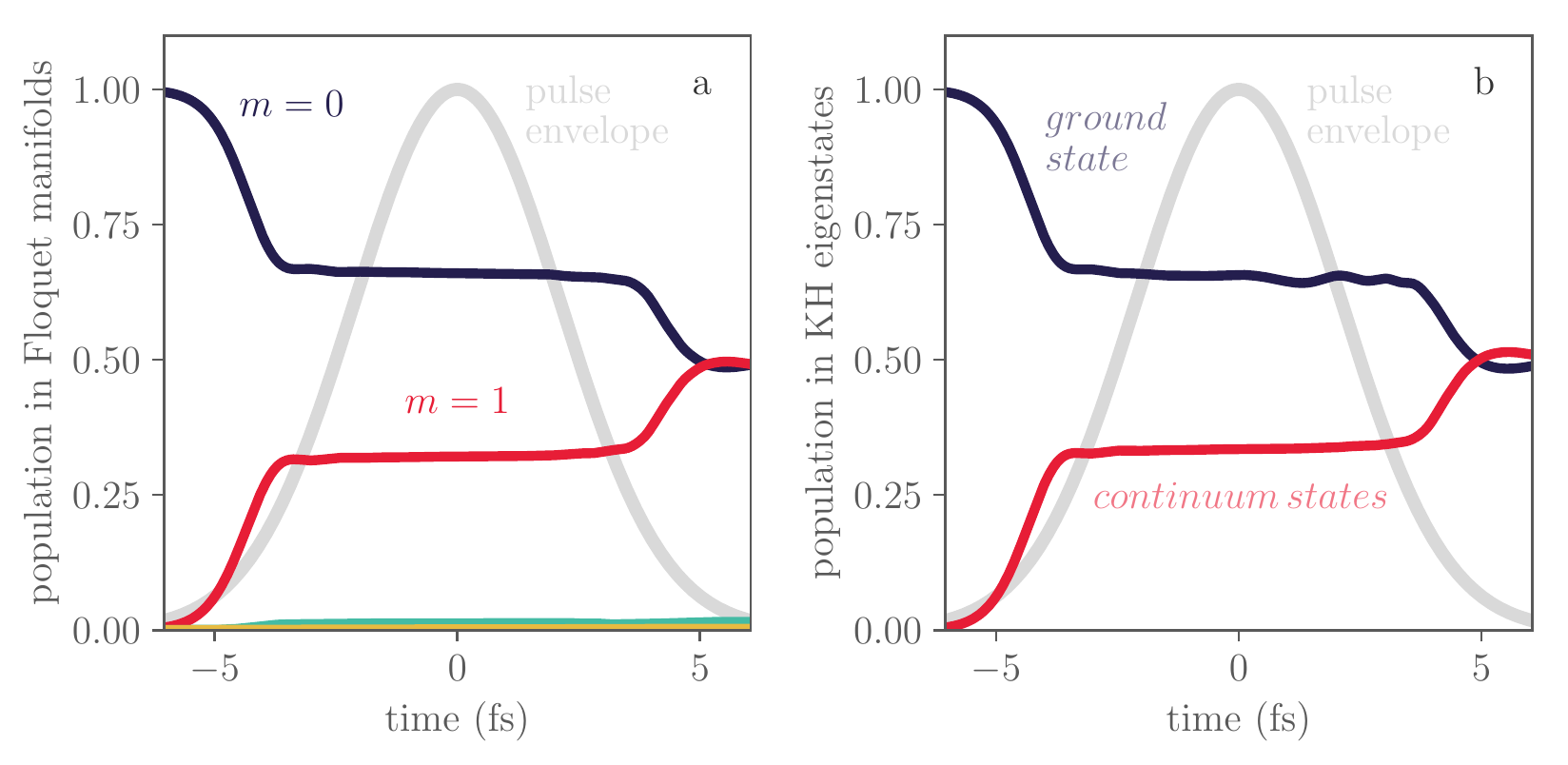}
	\caption{(a) Population in $m=0$ and $m=1$ Floquet channels as a function of time and
	(b) population in the ground state and the continuum states of the KH potential as a function of time for a laser pulse with frequency $\omega=1$ a.u., intensity $I=2.4 \times 10^{18} \, \text{W/cm}^2$ and duration $T=5$~fs FWHM pulse.}
	\label{fig:KH_and_Floquet_w1_5fs_a10}
\end{figure*}

%

The KH approach was originally proposed in the context of high-laser frequencies, for which the underlying dynamics is by now mostly well understood, see \cite{Gavrila2002} and \cite{Popov2003} for comprehensive reviews. Therefore, the high-frequency case provides a good reference point to illustrate the influence of the pulse envelope on the dynamics induced by high-intensity laser fields using the time-dependent Floquet approach developed here. We choose a laser frequency of $\omega = 1 \, \text{a.u.} \sim 27 \, \text{eV}$, substantially larger than the field-free ionization potential. Accordingly, the peak laser intensity is set to $I=2.4 \times 10^{18} \, \text{W/cm}^2$, so that the maximum electron excursion length is $\as_0(t=0)=10$ a.u. Note that for these laser parameters, non-dipole effects contribute negligibly to the dynamics \cite{Førre2005}. Hence, we safely work in the dipole approximation.

The population in the Floquet channels $m=0$ and $1$ as a function of time is depicted in \fig \ref{fig:KH_and_Floquet_w1_5fs_a10}a. During the initial part of the pulse around 30\% of the population is transferred from $m=0$ to the $m=1$ Floquet channel, indicating a one-photon absorption process. The population transfer stops, when the adiabatic stabilization regime is reached. Around the peak of the pulse, despite the rapid increase of field strength, population in each Floquet channel $m $ stays approximately constant, implying that the Floquet channels are decoupled as predicted by the high-frequency Floquet theory \cite{Marinescu1996}. As the field intensity decreases at the end of the pulse, another 20\% of the population is transferred to the $m=1$ Floquet channel by single photon absorption. The remaining Floquet channels $(m>1)$ contain $\ll 1$\% of the population after the pulse. 

\subsubsection{Non-adiabatic excitations}

\begin{figure}
\centering
\includegraphics{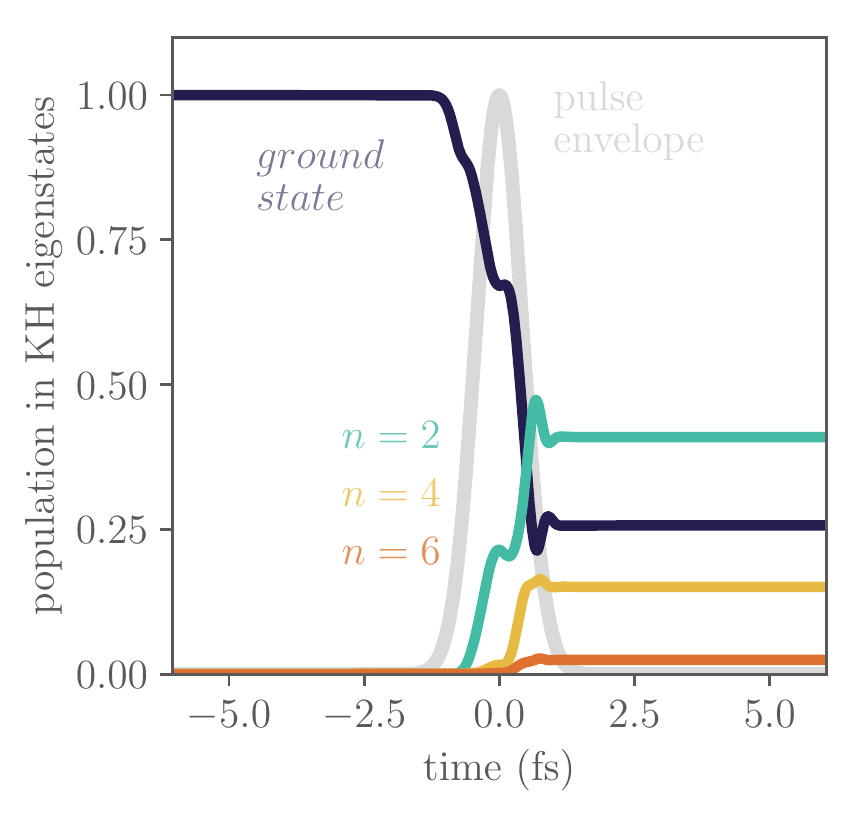}
\caption{Population in the ground state and $n=2, 4, 6$ excited states of the cycle-averaged potential in the $m=0$ Floquet channel as a function of time for the same pulse as in \fig \ref{fig:KH_and_Floquet_w1_5fs_a10} but with $T=1$~fs FWHM pulse.}
\label{fig:KH_w1_1fs_a10}
\end{figure}

Projecting the population in each Floquet channel $m$ onto the eigenstates of \CAHt\ reveals that the ionization process is adiabatic, see \fig \ref{fig:KH_and_Floquet_w1_5fs_a10}b.
The population in the $m=0$ subspace stays in the ground state throughout the dynamics and no substantial transitions due to the time-dependence of \CAHt\ takes place.
In this case, the ionization process can be described in terms of a discrete state that belongs to  the $m=0$ Floquet channel embedded into the continuum of states that belong to the $m=1$ channel, as assumed within the high-frequency approximation \cite{Marinescu1996}. Furthermore, once adiabatic stabilization sets in, the envelope of the laser pulse plays a minor role.

The adiabatic picture is not applicable for a shorter laser pulse with the same peak intensity. In this case, due to the rapid change of the eigenstates, non-adiabatic excitations from the ground to the excited states occur as is shown in \fig \ref{fig:KH_w1_1fs_a10} for a $T=1$~fs FWHM pulse. However, the population stays in the $m$=0 Floquet channels, i.e., no photons are absorbed from the field indicating that the excitations are induced by the envelope of the pulse. This is confirmed by excitations of even-parity states only, as absorption of a photon would lead to the excitation of odd-parity states. At the end of the pulse, the excited states of \CAHt\ transform to the corresponding field-free states. Such non-adiabatic transitions \cite{Toyota2009} were investigated in \cite{Toyota} using the envelope Hamiltonian formalism, where it was shown that they can be quantified using time-dependent perturbation theory.


\subsection{Intermediate-frequency} \label{sec:mid_freq}

\begin{figure}
	\centering
	\includegraphics[height=0.8\textheight]{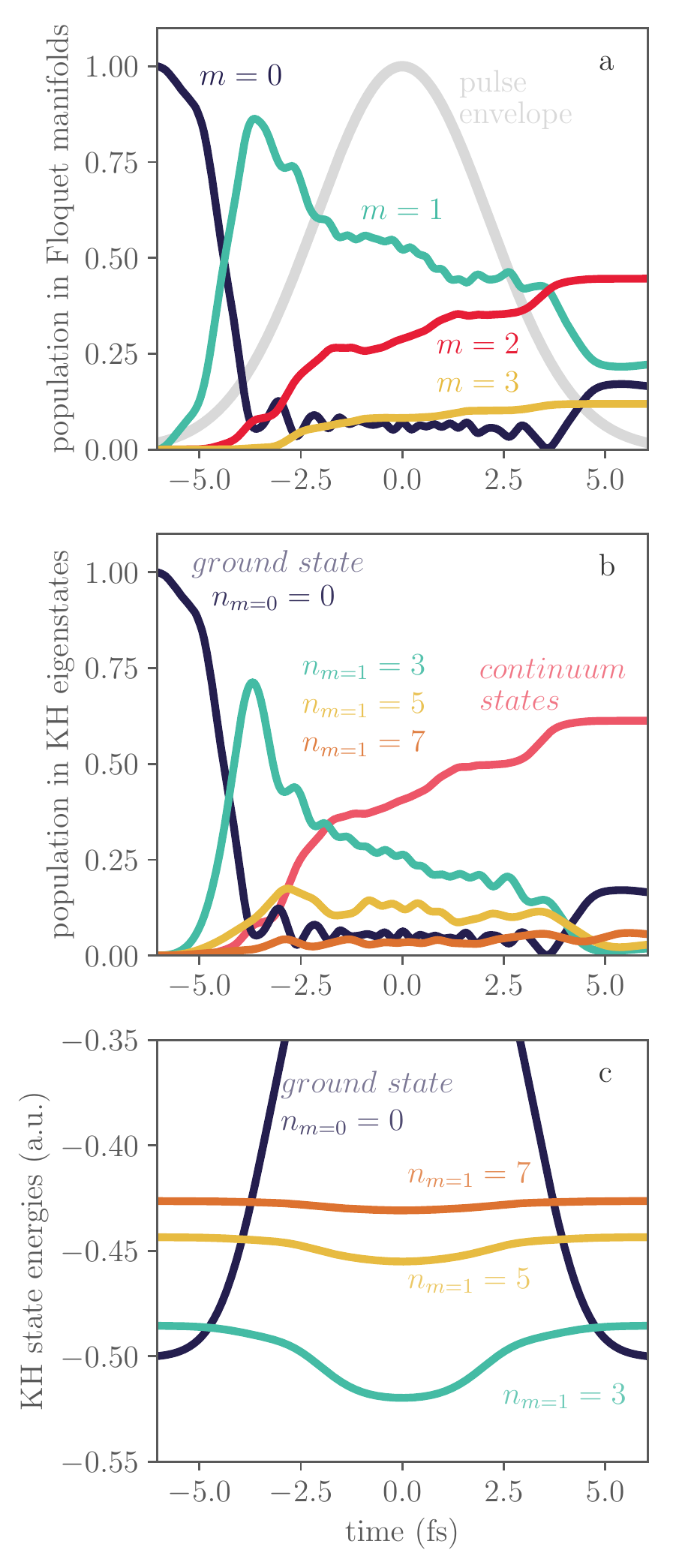}
	\caption{(a) Population in $m=0, 1$ and $m=2$ Floquet channels as a function of time; 
	(b) population in the ground state of $m=0$ Floquet channel, $n=3, 5, 7$ excited states of $m=1$ Floquet channel and continuum states of \CAHt\ as a function of time for $\omega=0.4$ a.u. frequency and $I=9 \times 10^{16} \, \text{W/cm}^2$ intensity laser pulse; (c) energies of the $n=3, 5, 7$ excited states from $m=1$ Floquet channel as a function of time, with the ground state energy of $m=0$ Floquet channel.}
	\label{fig:Triplet_w041_5fs_a10}
\end{figure}

%
%

A range of ``intermediate" frequencies can be defined, where the laser frequency is smaller than the binding energy of the field-free potential, but larger than the binding energy of the cycle-averaged potential at peak intensity. We will show, that for such ``intermediate" frequencies, the field-free ground state does not simply adiabatically connect to the ground state of \CAHt, as in the high-frequency case. Instead, it undergoes a series of crossings with excited states that belong to higher Floquet channels. 

We choose a laser frequency of $\omega = 0.4 \, \text{a.u.} \sim 10.9 \, \text{eV}$ and an intensity of $I= 9 \times 10^{16} \, \text{W/cm}^2$. Therefore, two photons are required for ionization, however the photon energy is still twice the binding energy of the cycle-averaged potential at the peak of the pulse, see \fig \ref{fig:Ladder}a.

The time-dependent populations in the $m=0, 1$ and $2$ Floquet channels, which are shown  in \fig \ref{fig:Triplet_w041_5fs_a10}a, immediately suggest that ionization proceeds in a sequential manner via the intermediate $m=1$ channel.
This is confirmed by the time-dependent population in the eigenstates of \CAHt\ depicted in \fig \ref{fig:Triplet_w041_5fs_a10}b. While the ground state of the $m=0$ Floquet channel is rapidly depopulated, the population is transferred to the odd-parity excited states of the $m=1$ channel, i.e., via one-photon transition. From these states, the population is slowly transferred to the $m=2$ channel, i.e., ionized via absorption of a further photon. 


\begin{figure*}
	\centering
	\includegraphics{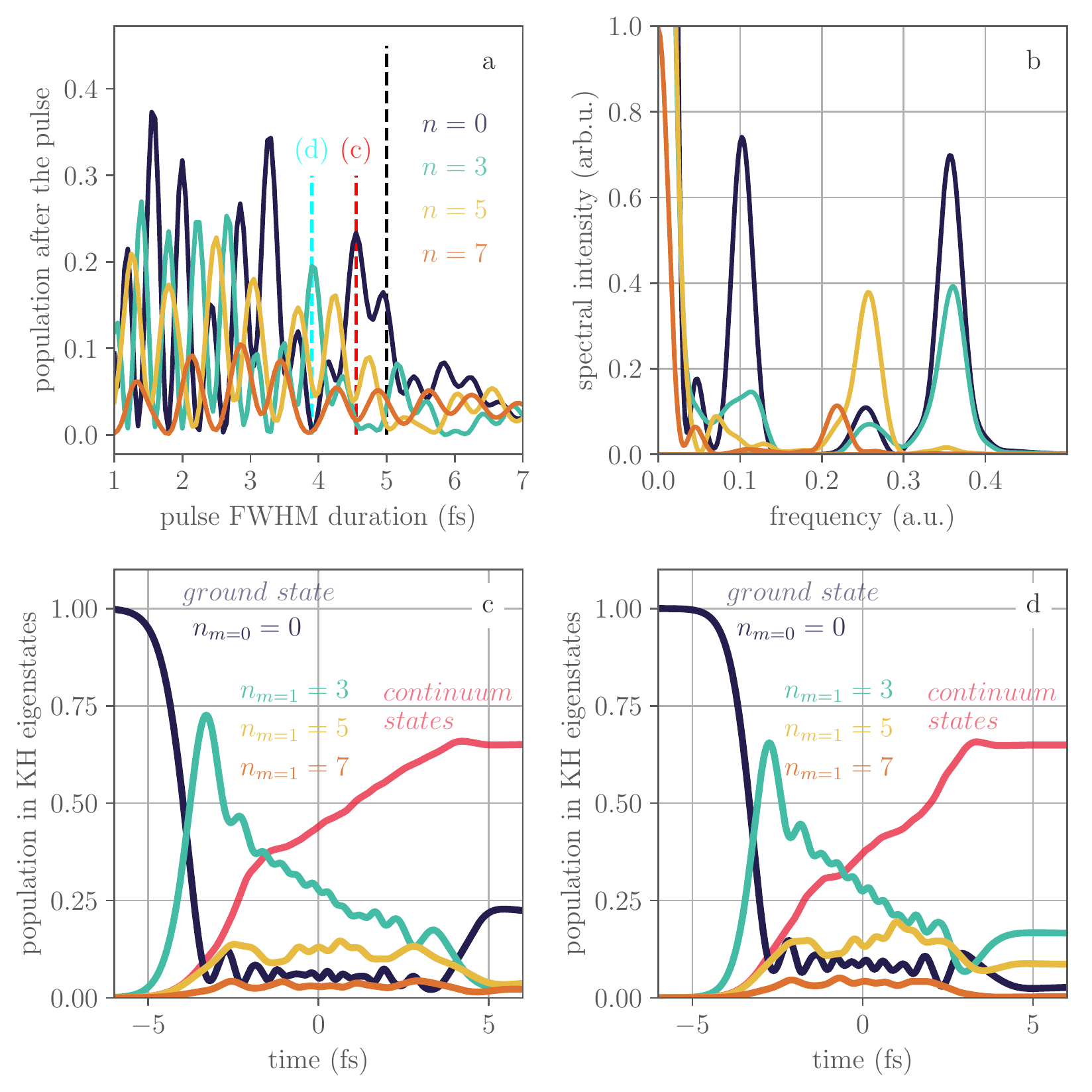}
	\caption{(a) Final population in the  ground state and $n=3, 5, 7$ excited states of \CAHt\ as a function of pulse duration for $\omega=0.4$ a.u. frequency and $I=9 \times 10^{16} \, \text{W/cm}^2$ maximum intensity laser pulse; the vertical dashed line indicates the pulse duration used in \fig \ref{fig:Triplet_w041_5fs_a10} and pulse durations when $n=0$ and $n=3$ state dominate; (b) frequency of population oscillations for each excited state;  (c) and (d) shows time-dependent population in the excited states for pulse durations when $n=0$ and $n=3$ state are dominantly populated after the pulse (indicated by vertical dashed lines in (a)).}
	\label{fig:KH_w041_scan_fwhm_a10}
\end{figure*}

The dynamics in \fig \ref{fig:Triplet_w041_5fs_a10}a and \ref{fig:Triplet_w041_5fs_a10}b can be understood in terms of the evolution of \CAHt\ eigenstates during the pulse, shown in \fig \ref{fig:Triplet_w041_5fs_a10}c.
At the beginning of the pulse the energy of the ground states rapidly increases and undergoes a series of crossings with the excited states that belong to the $m=1$ Floquet channel. At each of these crossings, a fraction of ground state population is transferred to the excited state. As the energy of the ground state decreases at the end of the pulse, the population is exchanged again at the second crossing. Between these crossings, a small but significant coupling of the states leads to small Rabi oscillations that are seen around $t=0$ in \fig \ref{fig:Triplet_w041_5fs_a10}b.


The two transitions that occur at the crossings of the ground and excited states of \CAHt\ lead to interference that depends on the phase accumulated in each state in between. This phase in turn depends on both the energy differences and the couplings between the states. Since the time between two crossings depends on the pulse duration it strongly influences the final population after the pulse, as seen in \fig \ref{fig:KH_w041_scan_fwhm_a10}a. 

\editt{
The origin of oscillations in \fig \ref{fig:KH_w041_scan_fwhm_a10}a is further elucidated by the evolution of excited-state populations during the pulse. In \fig \ref{fig:KH_w041_scan_fwhm_a10}c and \fig \ref{fig:KH_w041_scan_fwhm_a10}d these populations are shown for pulse durations, when either $n=0$ or $n=3$ state is predominantly populated after the pulse. Initially the dynamics in both cases is very similar. Clear differences emerge only after the second crossing between the states, indicating that interference effects determine the final populations. }

The final populations oscillate with well-defined frequencies as the pulse duration changes, see \fig \ref{fig:KH_w041_scan_fwhm_a10}b. 
The biggest amplitude oscillation is between the ground and $n=3$ state, which is to be expected since the coupling between these states is at least a factor of two larger than between any other states and $n=3$ state is the first one to undergo a crossing with the ground state. The time-window of strong interaction with the ground state is also longest for the $n=3$ state, since the crossing occurs at the beginning of the pulse, where the energy--time gradient is not as steep as for higher-energy states.

Unlike the final population of each state in \fig  \ref{fig:KH_w041_scan_fwhm_a10}a, which requires one to consider all interactions, we found that the frequencies of oscillation of final populations in \fig \ref{fig:KH_w041_scan_fwhm_a10}b are determined mainly by the dynamics of the ground and a single excited state.
The presence of higher-energy states does not significantly perturb these frequencies, since they mainly depend on  the phase difference accumulated between the times of crossing of the two states. 
\editt{
These times in turn depend on their energy difference -- the higher the energy of the excited state, the later the first crossing will occur. Higher-energy states contribute to smaller frequencies reducing their influence. 
}

Interaction of the ground state with any individual excited state can be readily described by a Landau-Zener-St\"{u}ckelberg (LZS) interference process \cite{Schafer1997a, Shevchenko2010}. In case of multiple states with non-trivial time-dependence of energies and couplings, the interconnected LZS transitions lead to complicated and rich dynamics. Nevertheless, characteristic features prevail. The ground state will be depleted sequentially transferring population to higher excited states at later times. Therefore later crossings will become less important due to weaker couplings and the smaller population available for transfer. Hence, the traces of single state dynamics show up in \fig \ref{fig:KH_w041_scan_fwhm_a10}b even for very high laser intensities.

\subsection{Low frequencies} \label{sec:low_freq}

\begin{figure}
\centering
\includegraphics{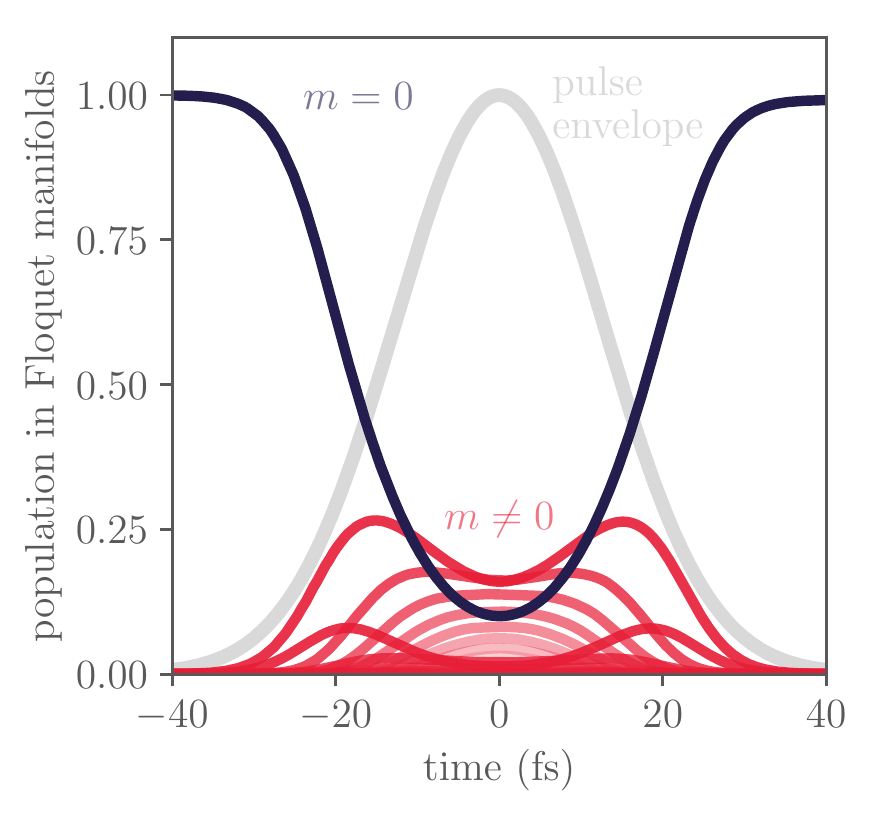}
\caption{Population in $|m| \leq 12$ Floquet channels as a function of time for $\omega=0.057$ a.u.  frequency and $I=3.7 \times 10^{13} \, \text{W/cm}^2$  intensity laser pulse. The gray line indicates the envelope of the laser pulse.}
\label{fig:Floquet_w005_30fs_a10}
\end{figure}

\begin{figure*}
\centering
\includegraphics{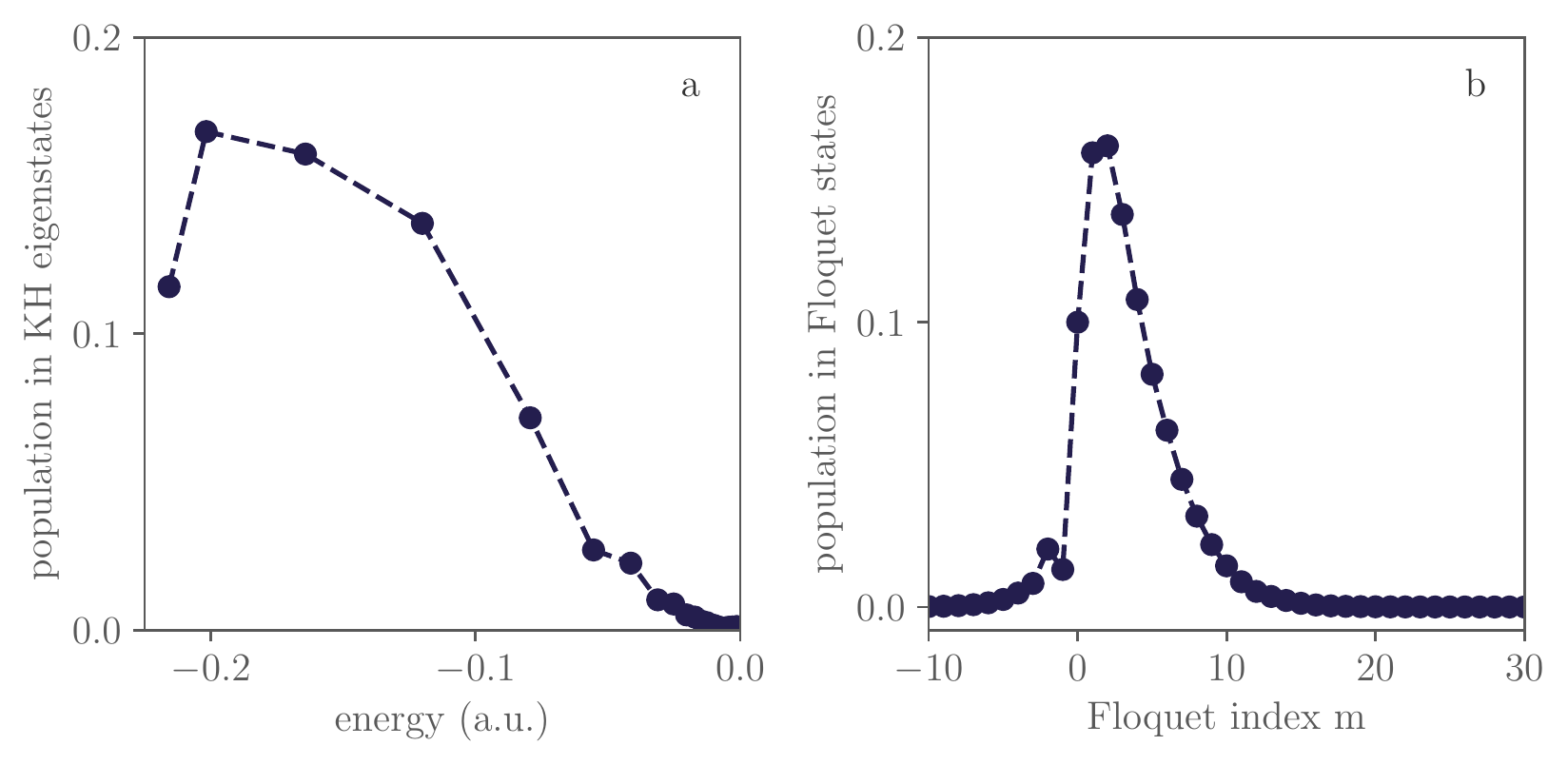}
\caption{Population in (a) the bound states of \CAHt\ and (b) in the Floquet channels at the peak of the laser pulse for $\omega=0.057$ a.u. frequency and $I=3.7 \times 10^{13} \, \text{W/cm}^2$ intensity laser pulse.}
\label{fig:low_freq_combined}
\end{figure*}

Although the KH approach was originally proposed to study the interaction of atoms with high-frequency laser fields, it was speculated that it could also be applicable for low-frequency and high intensity radiation \cite{Popov1999,Smirnova2000,Popov2011,Morales2011}. More recently, the KH approach and in particular the properties of the cycle-averaged potential was used to explain the nonlinear Kerr effect in laser filamentation \cite{Patchkovskii2013} and acceleration of neutral atoms in laser fields \cite{Wei2017}.
The time-dependent Floquet formalism developed here allows one to directly investigate the KH approach for frequencies that are much smaller than the ionization potential. 

We choose the laser frequency of $\omega=0.057$ a.u., which corresponds to $\lambda=800$ nm wavelength radiation, and $I=3.7 \times 10^{13}$W/cm$^2$ intensity, so that the maximal electron excursion length is again $\as_0=$10 a.u. Therefore, the eigenenergies of \CAHt\ and its dependence on the pulse shape shown in \fig \ref{fig:Ladder} is identical to the high and intermediate-frequency cases analyzed above. The FWHM duration of the pulses is set to $T=30$~fs. 
However, the essential results presented here do not depend sensitively on the duration of the pulse.
Note that in order to obtain converged results, 201 Floquet channels ($\pm 100 \, \omega$) are treated explicitly in the numerical calculation.

The population in Floquet channels as a function of time is plotted in \fig \ref{fig:Floquet_w005_30fs_a10}. 
Almost all of the population is transferred to higher Floquet channels at the peak of the pulse and then returns to the $m=0$ channel at the end of the pulse.
Therefore, these transitions are virtual, which is in contrast to high and intermediate-frequency cases. This is not surprising, however, since for the given laser field parameters the total ionization is less than $1\%$ and most of the population is expected to stay in the ground state after the pulse. 

Projecting the population at the peak of the pulse onto the instantaneous eigenstates of \CAHt\ in \fig \ref{fig:low_freq_combined}a
reveals that the population is distributed over many states. Crucially, no single state is dominating. 
Also, many Floquet channels are populated during the peak of the pulse, as is shown in \fig \ref{fig:low_freq_combined}b.
An increase of the field intensity leads to a broadening of the distribution over both the excited states $n $ and also over Floquet channels $m$.

Virtual excitations created in multiple Floquet channels can be understood qualitatively within the Floquet picture. Since the interaction strength between states that belong to different channels is much larger than the energy spacing between them, a quasi-continuum of states is created. 
In this situation, the eigenstates of \CAHt\ do not correspond to any adiabatic or nearly adiabatic states of the  field driven system. Therefore, the wave function, which stays nearly identical to that of the field-free ground state, is distributed over many excited states in the KH reference frame. 
The redistribution occurs at avoided crossings between states that belong to different Floquet channels, similarly as in the intermediate-frequency case. However, for low frequencies many more avoided crossings become important.

For sufficiently large peak field strengths a regime may exist, where \CAHt\ becomes applicable \cite{Popov1999,Smirnova2000,Popov2011}. However, by the time this intensity of the laser pulse is reached, the wave function is already distributed over the excited states of \CAHt. \hide{Therefore, the transformation of the field-free ground state to the ``field-dressed" KH picture, which is determined by the switching on of the laser pulse, is an essential step in order to apply the KH approach at low laser frequencies.}
\editt{Therefore, in order to apply the KH approach at low laser frequencies, it is essential to consider the transformation of the field-free ground state wave function to the ``field-dressed" KH picture during the switching-on of the pulse.}

\pagebreak
\section{Summary and conclusions}

We have developed a time-dependent Floquet approach formulated in the KH reference frame to study dynamics driven by short and intense laser pulses. It constitutes a systematic and flexible extension of the Envelope Hamiltonian \cite{Toyota} applicable for arbitrary frequencies and provide a convenient and efficient way to propagate Floquet Hamiltonians. 

\editt{Numerical application of Floquet approaches is often hampered by the rapid increase of the number of Fourier components required to describe the Hamiltonian.}
\hide{Numerical application of Floquet approaches is often hampered by the rapid growth of the Hamiltonian which requires an increasing number of Fourier components.} Therefore, we have devised an efficient numerical procedure to propagate the Floquet Hamiltonian which is able to overcome the hurdle of large expansions. Indeed, we have performed calculations for laser parameters, for which several hundred Floquet channels had to be considered explicitly. 
Key element is the formulation of the problem in the momentum representation, which is particularly suited for the KH reference frame as it allows us to separate the components of the field-free potential from the field-dependent ones.
We further use the formalism of Toeplitz matrices and the FFT algorithm to achieve a favorable scaling with Floquet channels. However, unlike the split-operator methods that also rely on FFT, the Toeplitz approach is numerically exact for finite-size matrices. For Floquet problems it allows us to truncate the basis to only several Floquet channels. Yet, the method can be applied to any other Fourier basis to achieve an efficient and accurate propagation.

The main advantage of the method is its ability to extend the KH approach, which is particularly suited for high-frequency and high-intensity fields, to the limit of very short pulses. Thereby, we can investigate physical effects that emerge at high intensities and can only be understood by explicitly considering the time-evolution of the pulse envelope. 

We have shown that the pulse envelope exerts control over two types of dynamics. For very short pulses, the rapid 
change of the eigenstates of \CAHt\ over time leads to non-adiabatic excitations. They are induced by the pulse envelope and therefore do not involve the absorption of any photons from the field. Thus, even for very high frequencies they can lead to significant population in low lying bound states. 

The second type of transitions, sensitive to the pulse envelope, occurs at the crossings between the discrete eigenstates of \CAHt\ that belong to different Floquet channels, as their energy changes along the pulse. 
Although dynamics at each crossing can be easily understood in terms of Landau-Zener-St\"{u}ckelberg theory, in our case multiple states are strongly coupled evading simple interpretations. Nevertheless, strong features due to the coupling between individual states can be discerned, which is quite remarkable, considering the high intensities used. They lead to a large sensitivity of final state populations to the pulse duration providing a possible route for their coherent control.

An extreme case for our KH approach is the low-frequency limit, when the photon energy is much smaller than the binding energy of the electron. In this case, the population is transferred between the bound states of \CAHt\ at their crossings, which are very dense due to the small energy spacing between the Floquet channels, resulting in a rapid distribution of the population over many eigenstates of \CAHt\ before any populations has a chance to reach continuum states.

To summarize, the time-dependent Floquet approach presented here provides a convenient basis for short laser pulses and for all but the smallest photon energies. In all investigated cases, including few-cycle pulses, the approach was able to provide accurate numerical results indistinguishable from the ones obtained using conventional techniques of propagating the TDSE. However, unlike the conventional TDSE propagators, the time-dependent Floquet method allows one to obtain insight into dynamics that crucially depend on the pulse envelope. Such dynamics  will become particularly important for short and intense pulses generated by FEL facilities, which often devise unusual pulse
shapes.

The richest envelope-dependent dynamics is observed in the intermediate-frequency range. For multi-electron systems, this energy range will be much more extended due to the ubiquitous presence of core and double excitations, which lead to a rich energy structure even for very high photon energies. Therefore, we expect that for such systems the time-dependent Floquet formalism presented here will be even more valuable.

%
%
%
%
\pagebreak
\appendix
\section{Derivation of the time-dependent Floquet formalism}\label{sec:app_expansions}

\subsection{Expansion of the Kramers--Henneberger potential into Fourier components}


In this work, the potential in the KH reference frame is expanded into Fourier and plane-wave components as
\begin{equation}
V(\rv, t) = \sum_{m} \intK \; \Vmk \, \mathrm{e}^{-\imath m \omega t} \, 
\e{\imath \kv \cdot \rv}. \label{eq:app_Fourier_expansion_V_a}
\end{equation} 
To determine the expansion coefficients $\Vmk$, let us first consider only the expansion into the Fourier components
\begin{equation}
V(\rv, t) = \sum_{m} \, \Vidx{m}(\rv, t) \, \e{-\imath m \omega t},\label{eq:app_Fourier_only_expansion_V} 
\end{equation}
which are determined by
\begin{equation}
\Vidx{m}(\rv, t) = \frac{1}{T_{\omega}} \, \int_0^{T_\omega} \, \dd t' \, \Vb(\rv, t, t') \, 
\e{\imath m \omega t'}.\label{eq:app_Fourier_expansion_V_b}
\end{equation}
The time-integration in Eq.~(\ref{eq:app_Fourier_expansion_V_b}) is performed only over $t'$, i.e., the periodic oscillation of the two-time potential
\begin{equation}
\Vb(\rv, t, t') = V\big(\rv + \av_0(t)\cos(\omega t'+\phase)\big).\label{eq:app_Fourier_expansion_V_c} 
\end{equation}
Nevertheless, this provides an \emph{exact} representation of the full time-dependent potential $V(\rv, t)$, as is easily verified 
by inserting Eq.~(\ref{eq:app_Fourier_expansion_V_b}) into Eq.~(\ref{eq:app_Fourier_only_expansion_V}):
\begin{multline}
V(\rv, t) =
\sum_m \Big( 
\frac{1}{T_\omega} \int_0^{T_\omega} \, \dd t' \, \Vb(\rv, t, t') \, \e{\imath m \omega t'}
\Big) \e{-\imath m \omega t} = \\
=\frac{1}{T_\omega} \int_0^{T_\omega} \, \dd t' \, \Vb(\rv, t, t') \, \sum_m \e{\imath m \omega (t'-t)} = \\
=\int_0^{T_\omega} \, \dd t' \, \Vb(\rv, t, t') \, \delta_\omega(t'-t) = V(\rv, t), \label{eq:V_tt_to_t}
\end{multline}
where the definition of the Dirac delta function
\begin{equation}
\delta_\omega(t'-t) = \frac{\omega}{2\pi} \sum_m \e{\imath m \omega (t'-t)} = 
\frac{1}{T_\omega} \sum_m \e{\imath m \omega (t'-t)} \label{eq:delta_def}
\end{equation}
was used.
Note that $\delta_\omega(t'-t)$ is periodic, with the period $T_\omega=2\pi/\omega$. However, its use is justified since $\Vb(\rv, t, t'+T_\omega)=\Vb(\rv, t, t')$ and the integration in Eq.~(\ref{eq:V_tt_to_t}) can be limited to the range $t' \in [0, T_\omega)$.

\subsection{Expansion of the Kramers--Henneberger potential into plane-wave components}

The Fourier components $\Vidx{m}(\rv, t)$ are further expanded into the basis of plane-waves. Using the definition of the potential in Eq.~(\ref{eq:app_Fourier_expansion_V_b}) and Eq.~(\ref{eq:app_Fourier_expansion_V_c}) the expansion coefficients can be written as (see also \cite{Yao1992} for a similar derivation)
\begin{subequations}
	\begin{align}
	\Vmk &= \KNORM \int \dd^3 r \, \Vidx{m}(\rv, t) \, \e{-\imath \kv \cdot \rv} \\
	&= \KNORM \frac{1}{T_\omega} \, \int_0^{T_\omega} \, \dd t' \, 
	\Big(\int \dd^3 r \, \Vb(\rv, t, t') \, \e{-\imath \kv \cdot \rv} \,\Big)
	\e{\imath m \omega t'} \\
	&=\KNORM \frac{1}{T_\omega} \, \int_0^{T_\omega} \, \dd t' \, 
	\Big(\int \dd^3 r \, V\big(\rv +\av(t, t')\big) \, \e{-\imath \kv \cdot \rv} \,\Big)
	\e{\imath m \omega t'} \\
	&= \frac{1}{T_\omega} \, \int_0^{T_\omega} \, \dd t' \, 
	\Vk \, \e{\imath \kv \cdot \av(t, t')} \,
	\e{\imath m \omega t'},
	\end{align}
\end{subequations}
where $\Vk$ is the projection of the field-free potential on the $\kv$-th plane-wave
\begin{equation}
\Vk = \KNORM \int_{-\infty}^{\infty} \dd^3 r' \, \mathrm{V}(\rv') \, \e{-\imath \kv \cdot \rv'},
\end{equation}
with $\rv' = \rv + \av(t, t')$.
Using
\begin{equation}
\av(t, t') = \av_0(t) \, \cos(\omega t' + \phase)
\end{equation}
and 
applying the Jacobi-Anger expression, the plane-wave Fourier components $\Vmk$ can be further expressed as
\begin{equation}
\Vmk = \Vk \, \imath^{|m|} \, J_{|m|}\big( \kv \cdot \av_0(t)\big) \, \e{-\imath m \phase},\label{eq:app_Vmk_equation} 
\end{equation}
where $J_m$ is the ordinary Bessel function of the 1st kind of order $m$.

\subsection{Derivation of the time-dependent Schr\"{o}dinger equation for the coupled Fourier and plane-wave components}

After expanding the wave function in terms of Fourier and plane-wave components
\begin{equation}
\Psi_{\khsub}(\rv, t) = \sum_{m} \intK \; \Pmk \, \e{-\imath m \omega t} \, \e{\imath \kv \cdot \rv} \label{eq:app_ansatz_wf_fourier}
\end{equation}
the expansion coefficients $\Pmk$ are determined by inserting Eqs.~(\ref{eq:app_Fourier_expansion_V_a}) and Eq.~(\ref{eq:app_ansatz_wf_fourier}) into the TDSE
\begin{equation}
\imath \frac{\partial}{\partial t} \Psi_\khsub(\rv, t) = -\frac{1}{2} \boldsymbol{\nabla} \Psi_\khsub(\rv, t) + V(\rv, t) \Psi_\khsub(\rv, t).
\end{equation}
After projecting on the $\kv$-th plane-wave component, the TDSE becomes
\begin{multline}
\sum_m \e{-\imath m \omega t} \Big( \imath \frac{\partial}{\partial t} \Pmk +m \omega \, \Pmk \Big) = \\
\sum_m \e{-\imath m \omega t} \Big( \frac{\kv^2}{2} \Pmk + \sum_{m'} \intK' \; \Vmkp \, \Pmkp \e{-\imath m' \omega t} \Big).
\end{multline}
Collecting the terms proportional to $\e{-\imath m \omega t}$ we obtain
\begin{multline}
\sum_m \e{-\imath m \omega t} \Bigg[  \Big(\frac{\kv^2}{2} -m \omega \Big) \Pmk + \sum_{m'} \intK' \; \Vmpkp \, \Pmpkp - \imath \frac{\partial}{\partial t} \Pmk \Bigg] = 0, \label{eq:collected_coefficients}
\end{multline}
where we have used that
\begin{multline}
\sum_{m,m'} \e{-\imath m \omega t} \e{-\imath m' \omega t} \, \Vmkp \, \Pmkp = \\
= \sum_{m''} \e{-\imath m'' \omega t}  \sum_{m} \Vt_{m''-m}(\kv-\kv', t) \, \Pmkp \\
= \sum_{m} \e{-\imath m \omega t} \sum_{m'} \Vmpkp \, \Pmpkp
\end{multline}
and where $m'' = m+m'$ was used in the second line together with relabeling of the indexes in the last line.
Eq.~(\ref{eq:collected_coefficients}) is satisfied, if the expression in brackets is zero for all times $t$ and for all $m$ and $\kv$. This condition leads to the coupled system of equations for the Fourier and plane-wave components of the wave function
\begin{equation}
\imath \frac{\partial}{\partial t} \Pmk  = \Big(\frac{\kv^2}{2} -m \omega \Big) \Pmk + \sum_{m'} \intK' \; \Vmpkp \, \Pmpkp. \label{eq:tdse_Fourier_components}
\end{equation}
Inserting the definition for $\Vmpkp$ from Eq.~(\ref{eq:app_Vmk_equation}) leads to the Eq.~(\ref{eq:tdse_tt}), which is the main equation used in this work.

\pagebreak

\section{Matrix-vector multiplication with Toeplitz and BTTB (Block Toeplitz with Toeplitz Blocks) matrices}

\subsection{Toeplitz matrix}\label{sec:app_T}

A matrix is Toeplitz if each of its diagonals is formed of equal elements. To describe a Toeplitz matrix only the knowledge of its 1st column and 1st row are required. Any $N \times N $ Toeplitz matrix can be  cast into a $2N \times 2N$ circulant matrix, which has identical rows and where each row is shifted to the right by one element with respect to the previous row, with rightmost element of the row transferred to the leftmost position. For example, a $3 \times 3$ Toeplitz matrix $\mathbf{T}$ can be transformed into a circulant matrix $\mathbf{C}$ as
\begin{equation}
\mathbf{T} =
\begin{bmatrix}
\mathsf{T}_{0} & \mathsf{T}_{-1} & \mathsf{T}_{-2} \\
\mathsf{T}_{1} & \mathsf{T}_{0} & \mathsf{T}_{-1}  \\
\mathsf{T}_{2} & \mathsf{T}_{1} & \mathsf{T}_{0} 
\end{bmatrix}
\Rightarrow
\mathbf{C} =
\begin{bmatrix}
\mathsf{T}_{0} & \mathsf{T}_{-1} & \mathsf{T}_{-2}  & 0 &  \mathsf{T}_{2} & \mathsf{T}_{1} \\
\mathsf{T}_{1} & \mathsf{T}_{0} & \mathsf{T}_{-1}  &  \mathsf{T}_{-2} & 0 & \mathsf{T}_{2} \\
\mathsf{T}_{2} & \mathsf{T}_{1} & \mathsf{T}_{0}  &  \mathsf{T}_{-1} & \mathsf{T}_{-2} & 0 \\
0 & \mathsf{T}_{2} & \mathsf{T}_{1} & \mathsf{T}_{0}  &  \mathsf{T}_{-1} & \mathsf{T}_{-2} \\
\mathsf{T}_{-2} & 0 & \mathsf{T}_{2} & \mathsf{T}_{1} & \mathsf{T}_{0}  &  \mathsf{T}_{-1} \\
\mathsf{T}_{-1} & \mathsf{T}_{-2} &  0 & \mathsf{T}_{2} & \mathsf{T}_{1} & \mathsf{T}_{0}  
\end{bmatrix} \label{eq:toeplitz_example},
\end{equation}
where zero was appended in each row to concatenate a Toeplitz matrix row with its column, and an arbitrary number of zeros can be used when forming a circulant matrix.

Multiplication of a circulant matrix $\mathbf{C}$ and a vector $\mathbf{\tilde{x}}$ is equal to a convolution between the first column of the circulant matrix $\mathbf{C}_{n0} \equiv \mathbf{c}$ and the vector $\mathbf{\tilde{x}}$. Hence, from the convolution theorem it follows that
\begin{align}
\mathbf{\tilde{b}} & = \mathbf{C} \cdot \mathbf{\tilde{x}} = \mathbf{c} \star \mathbf{\tilde{x}} \nonumber,\\
\mathcal{FT}(\mathbf{c} \star \mathbf{\tilde{x}}) &= \mathcal{FT}(\mathbf{c}) \, \mathcal{FT}(\mathbf{\tilde{x}}) = \mathcal{FT}(\mathbf{\tilde{b}}),
\end{align}
where $\star$ denotes the convolution operation and $\mathcal{FT}$ is the discrete Fourier transformation. 

To efficiently multiply a vector $\mathbf{x}$ by a general Toeplitz matrix $\mathbf{T}$ we need to: (i) form a column of the circulant matrix $\mathbf{c}$ from the 1st column and the 1st row of the Toeplitz matrix $\mathbf{T}$; (ii) append the vector $\mathbf{x}$ with zeros forming an extended vector $\mathbf{\tilde{x}}$ that has the same length as the vector $\mathbf{c}$; (iii) perform the Discrete Fourier Transformation of the vectors $\mathbf{c}$ and $\mathbf{\tilde{x}}$, multiply them together element-by-element and perform the inverse discrete Fourier transformation of the product. The result will be stored in the first $N$ elements, where $N$ is the size of the original vector. For the example in Eq.~(\ref{eq:toeplitz_example}):
\begin{align}
&\mathbf{b} = \mathbf{T} \cdot \mathbf{x} \Rightarrow 
\mathbf{\tilde{b}} = \mathbf{c} \cdot \mathbf{\tilde{x}}, \nonumber \\
\intertext{where} \nonumber\\
&\mathbf{T} \Rightarrow
\mathbf{c} = 
\begin{bmatrix}
\mathsf{T}_{0} \\
\mathsf{T}_{1} \\
\mathsf{T}_{2} \\
0 \\
\mathsf{T}_{-2} \\
\mathsf{T}_{-1}
\end{bmatrix},
\hspace{20pt}
\mathbf{x}=
\begin{bmatrix}
X_{0} \\
X_{1} \\
X_{2}
\end{bmatrix}
\Rightarrow
\mathbf{\tilde{x}}=
\begin{bmatrix}
X_{0} \\
X_{1} \\
X_{2} \\
0 \\
0 \\
0 \\
\end{bmatrix}, \nonumber \\ \nonumber \\
\intertext{and} \nonumber\\
&\mathbf{\tilde{b}} =  \mathcal{FT}^{-1}\Big(  \mathcal{FT}(\mathbf{c}) \, \mathcal{FT}(\mathbf{\tilde{x}}) \Big), \nonumber \\
&\mathbf{b}_i = \mathbf{\tilde{b}}_i \; \text{ for } i < 3.
\end{align}

\subsection{BTTB matrix}\label{sec:app_BTTB}

A Block Toeplitz with Toeplitz Block, or BTTB matrix, is a block matrix where the blocks are of Toeplitz form and the blocks on each diagonal are identical. For example, assume a BTTB matrix $\mathbf{B}$ with $3 \times 3$ blocks, where each block is a Toeplitz matrix $\mathbf{T}_m$ of the from in Eq.~(\ref{eq:toeplitz_example})
\begin{equation}
\mathbf{B} = 
\begin{bmatrix}
\mathbf{T}_{0} & \mathbf{T}_{-1} & \mathbf{T}_{-2} \\
\mathbf{T}_{1} & \mathbf{T}_{0} & \mathbf{T}_{-1}  \\
\mathbf{T}_{2} & \mathbf{T}_{1} & \mathbf{T}_{0} 
\end{bmatrix}
\text{  with  }
\mathbf{T}_m = 
\begin{bmatrix}
\mathsf{T}_{m,0} & \mathsf{T}_{m,-1} & \mathsf{T}_{m,-2} \\
\mathsf{T}_{m,1} & \mathsf{T}_{m,0} & \mathsf{T}_{m,-1}  \\
\mathsf{T}_{m,2} & \mathsf{T}_{m,1} & \mathsf{T}_{m,0} 
\end{bmatrix} \label{eq:example_bttb}.
\end{equation}

To calculate the dot product of a BTTB matrix $\mathbf{B}$ and a vector $\mathbf{x}$ one has to (i) form a 2D circulant matrix $\mathbf{C}$, where each column is the circulant vector for a Toeplitz block $\mathbf{T}_m$; (ii) reshape the vector $\mathbf{x}$ into a matrix $\mathbf{\tilde{X}}$ that has the same shape as the circulant matrix $\mathbf{C}$ by filling the lower half and the right half of $\mathbf{\tilde{X}}$ with zeros; (iii) perform the 2D Discrete Fourier Transformation of the matrices $\mathbf{C}$ and $\mathbf{\tilde{X}}$, multiply them together element-by-element and perform inverse 2D Discrete Fourier Transformation of the product. The result will be stored in the upper left corner of size $N \times M$, where $N$ is the size of each Toeplitz block $\mathbf{T}_m$ and $M$ is the number of block. For the example in Eq.~(\ref{eq:example_bttb}) the procedure is
\begin{align}
&\mathbf{d} = \mathbf{B} \cdot \mathbf{x} \Rightarrow \mathbf{\tilde{D}} 
= \mathbf{C} \cdot \mathbf{\tilde{X}}, \nonumber
\end{align}
where
\begin{align}
&\mathbf{B} \Rightarrow
\mathbf{C} = 
\begin{bmatrix}
\mathbf{c}_{0} & \mathbf{c}_{1} & \mathbf{c}_{2} & \mathbf{0} & \mathbf{c}_{-2} & \mathbf{c}_{-1}
\end{bmatrix}
= \nonumber \\
& =\begin{bmatrix}
\mathsf{T}_{0, 0} & \mathsf{T}_{1, 0} & \mathsf{T}_{2, 0} & 0 & \mathsf{T}_{-2, 0} & \mathsf{T}_{-1, 0} \\
\mathsf{T}_{0, 1} & \mathsf{T}_{1, 1} & \mathsf{T}_{2, 1} & 0 & \mathsf{T}_{-2, 1} & \mathsf{T}_{-1, 1} \\
\mathsf{T}_{0, 2} & \mathsf{T}_{1, 2} & \mathsf{T}_{2, 2} & 0 & \mathsf{T}_{-2, 2} & \mathsf{T}_{-1, 2} \\
0 & 0 & 0 & 0 & 0 & 0 \\
\mathsf{T}_{0, -2} & \mathsf{T}_{1, -2} & \mathsf{T}_{2, -2} & 0 & \mathsf{T}_{-2, -2} & \mathsf{T}_{-1, -2} \\
\mathsf{T}_{0, -1} & \mathsf{T}_{1, -1} & \mathsf{T}_{2, -1} & 0 & \mathsf{T}_{-2, -1} & \mathsf{T}_{-1, -1} 
\end{bmatrix}, \nonumber \\ \nonumber \\
&\mathbf{x}=
\begin{bmatrix}
X_{0} \\ X_{1} \\ X_{2} \\ X_{3} \\ X_{4} \\
X_{5} \\ X_{6} \\ X_{7} \\ X_{8} 
\end{bmatrix}
\Rightarrow
\mathbf{\tilde{X}}=
\begin{bmatrix}
X_{0} & X_{3} & X_{6} & 0 & 0 & 0 \\
X_{1} & X_{4} & X_{7} & 0 & 0 & 0 \\
X_{2} & X_{5} & X_{8} & 0 & 0 & 0 \\
0 & 0 & 0 & 0 & 0 & 0 \\
0 & 0 & 0 & 0 & 0 & 0 \\
0 & 0 & 0 & 0 & 0 & 0 
\end{bmatrix}, \nonumber \\ \nonumber
\end{align}
and
\begin{align}
&\mathbf{\tilde{D}} = \mathcal{FT}^{-1}_{2D}\Big(  \mathcal{FT}_{2D}(\mathbf{C}) \, \mathcal{FT}_{2D}(\mathbf{\tilde{X}}) \Big), \nonumber \\
& \mathbf{d}_{i+3j} = \mathbf{\tilde{D}}_{i,j} \; \text{ for } i,j < 3. 
\end{align}

\pagebreak
\section*{References}
\bibliographystyle{ieeetr}
\bibliography{bibliography}
\end{document}